\documentclass[aps,pre,onecolumn]{revtex4}
\usepackage{graphicx, bm, bbm,amsmath, amssymb,amsfonts, epsfig,color}
\usepackage{float}
\usepackage{todonotes}
\usepackage{xcolor}

\pdfoutput=1

\newcommand{\bq}{\begin{align}}
\newcommand{\eq}{\end{align}}

\usepackage{comment}

\begin{document}

\title{Wavy optimal flows for heat transfer in channels}

\author{Shivani Prabala}
\email[Electronic address: ]{sprabala@umich.edu}

\author{Silas Alben}
\email[Electronic address: ]{alben@umich.edu}

%\thanks{}
\affiliation{Department of Mathematics, University of Michigan,
Ann Arbor, MI 48109, USA}


\begin{abstract}
We compute incompressible two-dimensional fluid flows that maximize the rate of heat transfer from the walls of a straight channel given a specified flow input power $Pe^{2}$, where $Pe$ is the P\'{e}clet number. We use the Broyden–Fletcher–Goldfarb–Shanno (BFGS) algorithm together with an adjoint method to compute gradients. The optimal flows are approximately unidirectional up to a critical $Pe \approx 2^{12}$. Above this value the flows assume wavy patterns characterized by finger-like protrusions emanating from both the top and bottom walls of the channel. The rate of heat transfer for these wavy flows is 3\% to 30\% greater than that of the previously identified unidirectional optima \cite{Alben2017Channel} for $2^{13} \leq Pe \leq 2^{17}$. The wavy flows have a much smaller flux through the channel than the unidirectional flows, with regions of slow-moving fluid at nearly homogeneous temperature interspersed with serpentine regions of fast-moving fluid. Consequently, the area of the interface between hot and cold fluid is increased.
\end{abstract}
\maketitle

\section{Introduction}
\label{SectionI}
As energy efficiency becomes increasingly important, the study of heat transfer mechanisms continues to be a vital research area in a variety of engineering and industrial domains. Heat transfer is the process of moving thermal energy from one physical system to another. We focus on the classical mechanisms of conduction and convection \cite{Incropera2007}. Applications of heat transfer span numerous fields such as thermal management in electronics, energy efficiency in buildings, thermal power generation, cooking and food processing, aerospace engineering, climate control, environmental studies, and industrial manufacturing processes \cite{yovanovich2007advances, dahiya2019heat, sharif2018thermal, tsao2012heat, chung2014heat, zhang2016heat, patel2011heat}. Heat transfer enhancement encompasses methods that increase the amount of heat transferred in a system. Since the mid-20th century, researchers have explored a wide variety of methods for enhancement which have been classified broadly as active or passive techniques \cite{webb2005principles}. Passive techniques exploit modified surface geometries or fluid additives, while active techniques employ external power such as electric or acoustic fields and surface vibration \cite{bergles1983bibliography}.

Common passive strategies for enhancing convective heat transfer can be broadly categorized into geometric modifications, flow manipulation techniques, and material-based methods \cite{webb2005principles}. Geometric modifications include roughened or extended surfaces and coiled tubes. Roughened surfaces use features such as angled ribs, grooves, or vortex generators to disrupt the boundary layer and promote turbulence \cite{GEE19801127, PROMVONGE20101242, Han1985, TSIA19992071}. Extended surfaces such as fins, spines, or metal tapes increase the effective heat transfer surface area, thereby improving heat dissipation \cite{kraus2001extended}. Coiled tubes enhance heat transfer by inducing centrifugal forces that create secondary flows and increase turbulence, leading to higher local heat transfer coefficients \cite{kruthiventi2018coiled}. Flow manipulation techniques involve devices that alter the flow path or fluid velocity field to promote mixing. Displacement enhancement devices, such as wire mesh inserts or porous structures, increase turbulence by modifying the flow near the wall without directly altering the heated surface \cite{bozzoli2021characterisation}. Swirl flow generators, including twisted tapes, helical coils, and wire inserts, introduce rotational or helical motion, breaking the thermal boundary layer and enhancing convective transport \cite{YILMAZ19991365}. Material-based methods focus on modifying fluid or surface properties to improve thermal performance. Surface-tension-based strategies, such as microchannel structures, grooved surfaces, or wick materials, enhance capillary-driven flow and improve thermal contact. These are particularly effective in applications like heat pipes and micro-scale devices, where liquid spreading and thin-film evaporation are dominant mechanisms \cite{lin2021enhancement, xi2022surface}. Additionally, additives including nanoparticles, surfactants, or multiphase agents (e.g. bubbles or droplets) can modify fluid properties such as viscosity, specific heat, and thermal conductivity, thereby improving convective heat transfer characteristics \cite{saha2019electric}.

Active enhancement techniques can be broadly categorized into mechanical motion techniques, electric, acoustic, and vibrational field methods, and flow injection or extraction strategies \cite{bergles1983bibliography}. Mechanical motion techniques use moving surfaces or components to disrupt the thermal boundary layer and intensify mixing. Examples include scraped surface heat exchangers, rotating cylinders, and vibrating plates, which generate turbulence and enhance heat transfer rates \cite{blasiak2019numerical}. Electric, acoustic, and vibrational field methods apply external energy sources to induce fluid motion or disrupt boundary layers. Surface and fluid vibrations, introduced via acoustic, ultrasonic, or flow-induced mechanisms, increase turbulence, especially in single-phase flows \cite{al2006effect, rahman2020characterization, jadhav2016review}. Electrostatic field techniques use electric fields to drive fluid motion, reduce interfacial thermal resistance, or influence phase-change processes such as boiling and condensation \cite{CHEN2022104027, OZKAN2020119207, BOLOGA1995175}. A notable example is ion injection, an electrohydrodynamic (EHD) method where ions emitted from a high-voltage source generate localized cooling plumes that significantly enhance heat transfer \cite{testi2007ion}. Wall deformation techniques, such as synthetic jets, pulsatile flows generated by oscillating diaphragms, membrane-driven acoustic waves, and large-amplitude surface oscillations, also promote convective enhancement through induced mixing and boundary layer disruption \cite{LEAL2013505}. Flow injection and extraction strategies manipulate the flow near heated surfaces to improve thermal performance. Jet impingement, including spray cooling, directs high-velocity fluid jets at a surface, breaking the thermal boundary layer and increasing local turbulence. In spray systems, additional effects like evaporation and secondary nucleation further boost heat transfer. Suction, on the other hand, removes fluid from the boundary layer through porous or perforated surfaces, reducing thermal resistance and enhancing cooling efficiency especially under laminar flow conditions or during vapor removal in boiling regimes \cite{JENG199525, KHEZERLOO2021105377}.

Heat transfer for flow through a channel with heated walls is a well-known problem with early theoretical work by Graetz \cite{Graetz}.
In 1928, Lévêque developed a similarity solution for heat transfer in laminar flow with thin thermal boundary layers. His analysis led to what is now known as the Lévêque approximation, which relates the Nusselt number to the Reynolds and Prandtl numbers, showing that $Nu \propto (Re \cdot Pr)^{1/3}$. This solution remains fundamental in the study of convective heat transfer, particularly for high Péclet number flows \cite{leveque1928laws}.
Other solutions for heat transfer in channels can be found in the comprehensive reference by Shah and London, which provides analytical data and methods for laminar flow forced convection in ducts, particularly relevant to compact heat exchanger applications \cite{shah2014laminar}. Alongside experimental and analytical methods, computational fluid dynamics has been used to explore optimized channel flow designs for improved heat transfer. Simulations show that surface protrusions such as ribs or baffles can significantly enhance thermal transport while balancing pressure drop, with orthogonal rib configurations performing especially well \cite{ghobadi2021optimization, DU2025105850}. 

While much research focuses on geometric manipulation, another approach keeps the geometry fixed and optimizes the flow field itself. Hassanzadeh et al. formulated the problem of maximizing steady, incompressible heat transport between two parallel walls by optimizing the velocity field subject to constraints on either the kinetic energy or enstrophy of the flow, the ``wall-to-wall" problem \cite{hassanzadeh2014wall}. Using variational methods combined with numerical optimization, they identified optimal flow structures characterized by arrays of convection cells whose aspect ratios vary with flow intensity, and showed how $Nu$ scales with Pe, the P\'{e}clet number.

Tobasco and Doering developed a theoretical variational framework to construct steady two-dimensional incompressible flows that nearly achieve the theoretical upper bound on heat transport (in the context of Rayleigh-Bénard convection) under an enstrophy constraint. They demonstrated that the Nusselt number scales as $Nu \sim Pe^{2/3}$, up to possible logarithmic corrections \cite{tobasco2017optimal}. Alben employed constrained optimization to identify steady two-dimensional flows that maximize heat transfer in various geometries, including channels with heated top and bottom walls under fixed energy input \cite{alben2017geometries, Alben2017Channel}. Using Newton's method with continuation, he discovered optimal flow structures characterized by sharp boundary layers, enhanced thermal gradients, and nearly uniform velocity in the channel interior. For the wall-to-wall problem, Souza et al. \cite{souza2020wall} used a gradient ascent method and an asymptotic analysis to study optimal transport with no-slip rather than free-slip boundary conditions as in \cite{hassanzadeh2014wall}. Alben extended the wall-to-wall study to higher $Pe$ and found that branched flow structures emerged at a critical Pe, followed by multiple levels of branching  \cite{alben2023transition}, similar to an earlier 3D study \cite{motoki2018maximal}. In this paper, we extend these optimization studies to determine the flows that are optimal for heat transfer in channels with inflow and outflow. Like \cite{alben2023transition}, we reformulate the constrained optimization problem as an unconstrained one in a space of flow mode coefficients. We classify and analyze the resulting optimal flows and compare them to previous optima found in \cite{Alben2017Channel}. We find a transition from unidirectional to wavy flow structures beyond a critical $Pe$.
\section{Model}
\label{SectionII}
\begin{figure}
    \hspace{-0.1 in} 
    \includegraphics[width=7 in]{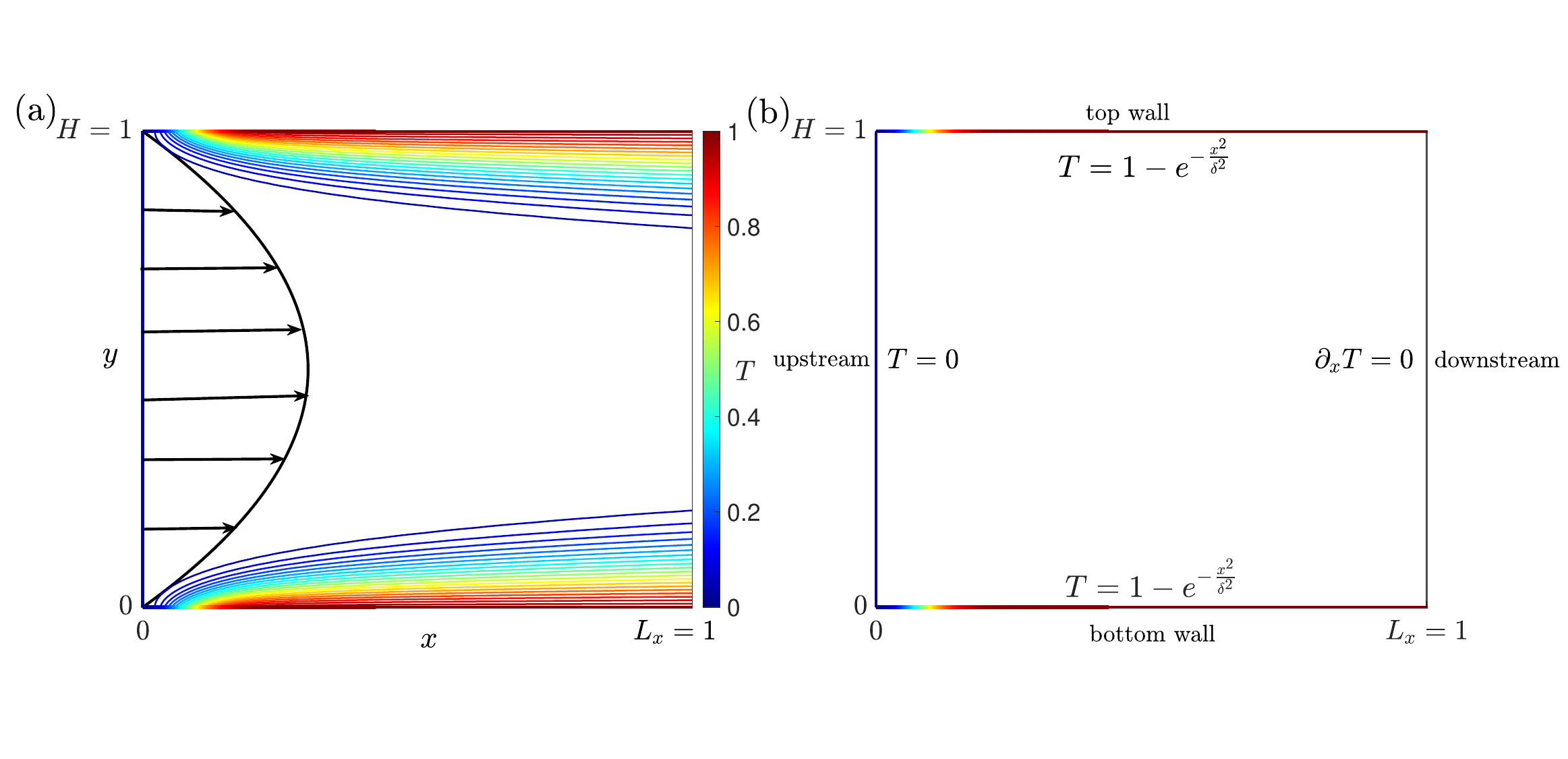}
    \vspace{-0.5 in}
    \caption{An example of the computational domain showing a parabolic flow through the channel, and contours of the resulting temperature solution. In panel (a), cold fluid (with dimensionless temperature $T$ = 0) enters the channel at the left, and flows past the heated top and bottom channel walls (with dimensionless temperature given by $ T_{\text{boundary}}(x, y) = 1 - e^{-\frac{x^{2}}{ \delta^{2}}}$). The dimensionless channel height and length are both 1. Panel (b) shows the temperature boundary conditions.}
\label{fig:ProblemSetUp_BC}
\end{figure}

We compute steady, incompressible two-dimensional fluid flows $\boldsymbol{u} = (u, v)$ that maximize the heat flux from the heated walls of a straight channel with height $H$ equal to its length $L_{x}$. The fluid domain is shown in figure~\ref{fig:ProblemSetUp_BC} and has $(x,y) \in [0,1]^2$ in dimensionless coordinates we now describe.
\subsection{Governing equations and boundary conditions}
\label{SectionIIA}

The fluid temperature field $T(x, y)$ satisfies the steady advection-diffusion equation: 
\begin{equation}
    \boldsymbol{u} \cdot \nabla T - \kappa \nabla^2 T = 0
    \label{eqn: adv-diff}
\end{equation}
where $\kappa$ is the thermal diffusivity of the fluid. The dimensional temperature along the top and bottom walls is $T_{\text{boundary}}(x, y) = T_{l} \left( 1 - e^{-\frac{x^{2}}{ \delta^{2}}}\right)$ where $\delta$ is a small smoothing parameter used to avoid a discontinuous temperature where the top and bottom walls meet the inflow boundary. For $x$ beyond the range $0 \leq x \lessapprox \delta$,
we have $T_{\text{boundary}} \approx T_{l}$.
We set $\kappa$, the dimensional thermal diffusivity of the fluid, to unity corresponding to nondimensionalizing velocities by $\kappa/L_x$, and we nondimensionalize lengths by $L_{x}$. 
We use the following set of dimensionless variables:  
\begin{equation}
    u^{*} = \frac{L_{x}}{\kappa} u, \hspace{0.2 in} v^{*} = \frac{L_{x}}{\kappa} v,\hspace{0.2 in} x^{*} = \frac{x}{L_{x}}, \hspace{0.2 in} y^{*} = \frac{y}{L_{x}}, \hspace{0.1 in} \text{and} \hspace{0.1 in}  T^{*} = \frac{T - T_{\text{inlet}}}{T_{l} - T_{\text{inlet}}}.
\end{equation}
Plugging these into (\ref{eqn: adv-diff}), we obtain the dimensionless form of the steady advection-diffusion equation: 
\begin{equation} \label{eqn:nondimadvecdiff}
    \boldsymbol{u^{*}} \cdot \nabla T^{*} - \nabla^{2} T^{*} = 0
\end{equation}
Henceforth we use dimensionless variables only and drop the $^*$s for convenience.
 At the inflow boundary, $x = 0$ and cold fluid enters with temperature $T = 0$. On the top and bottom walls ($y = 0$ and $y = 1$ respectively), the dimensionless temperature is $T = 1 - e^{-\frac{x^{2}}{\delta^{2}}}$. We set $\delta = 0.1$ so the smoothing occurs over a small region. At the channel exit, $x = 1$, we use the standard outflow condition $\partial_x T = 0$ \cite{shoele2014computational}.
 This condition implies that heat passes though the downstream boundary only by convection, not conduction.

 \subsection{Objective function for enhanced heat transfer}
 \label{SectionIIB}
 We maximize the Nusselt number, $Nu$, the dimensionless rate of heat transfer from the top and bottom walls of the channel, over the space of steady incompressible flows. $Nu$ is given by: 
\begin{equation} \label{eqn:Nu}
     Nu =  \int_{0}^{1} -\partial_y T  \Big|_{y = 0} + \partial_y T \Big|_{y = 1} dx.
 \end{equation}
We apply the optimization routine to flows that are a sum of two parts, a flow through the channel and an interior flow that does not pass through the upstream
and downstream boundaries. For the flow through the channel, we write: 
\begin{equation}
    \psi_{1}(x, y) = \psi_{\text{top,min}} (3y^{2} - 2y^{3})
\end{equation}
where $\psi_{\text{top,min}}$ is a positive constant that controls the magnitude of this Poiseuille flow. $\psi_{\text{top,min}}$ is chosen large enough to ensure a nontrivial flow through the channel, making the boundary conditions at the inlet and outlet physically valid. The interior flow component is essentially arbitrary and can represent internal vortices and recirculation zones for example, which may enhance mixing and convective transport. The interior flow is defined as:
\begin{equation} \label{eqn:Psi2}
    \psi_{2}(x,y) = \Big[\sum_{j = 1}^{M_2}\sum_{k = 1}^{N_2} B_{jk} P_j(x)Q_k(y) \Big] \cdot \Big(1 - e^{-\frac{x^{4}}{\delta^{4}}} \Big)
\end{equation} 
where $P_{j}(x)$ are linear combinations of Chebyshev polynomials that are zero at $x$ = 0 and 1, and have zero $x$-derivative at the upstream boundary. This ensures that $\psi_2$ vanishes at the upstream and downstream boundaries and prevents the optimization routine from converging to flow solutions that have strong eddies near the inflow boundaries which would alter the inflow temperature. The functions $Q_k(y)$ are linear combinations of Chebyshev polynomials that are zero and have zero $y$-derivative at $y = 0$ and $1$, thereby satisfying the no-slip boundary condition on the walls. The construction of the $Q_k(y)$ is described in appendix B of \cite{alben2023transition} and the construction of the $P_{j}(x)$ is similar. $B_{jk}$ are the coefficients for each basis function pair, giving the amplitude of the contribution from each mode. The exponential factor $(1 - e^{-\frac{x^{4}}{\delta^{4}}})$ smoothly suppresses this interior flow near the upstream boundary to further prevent inflow cancellation or distortion. In particular, this ensures that we find solutions that have a horizontal flow through the channel at the inlet, allowing us to find flows where the heat conduction is very small at the upstream boundary, so it is reasonable to assume $T$ remains zero there. We will illustrate the effects of applying this exponential factor in section \ref{SectionIVB4}, 
figure~\ref{EffectofSmallFlux}. The total stream function is then: 
\begin{equation}
    \psi(x, y) = \psi_{1}(x, y) + \psi_{2}(x,y).
\end{equation}
The combined flow will be normalized so that the rate of viscous energy dissipation, i.e. the power, is fixed at $Pe^2$, where $Pe$ is the P\'{e}clet number, the strength of advection relative to diffusion of heat in the flow. In the absence of bulk flow, heat transfer occurs solely by conduction, which is relatively inefficient \cite{Cengel9th}. By designing flows that enhance convective transport, via both a base flow and internal recirculation, we seek to significantly increase the Nusselt number and identify flow fields that optimize heat transfer performance. 
\subsection{Energy dissipation in viscous flow}
\label{SectionIIC}
The rate of energy dissipation of a viscous flow is
\begin{align}
   \mbox{Power} = 2\mu W \iint e_{ij}^2 dx dy,
\end{align}
with the integral taken over the flow domain, $\mu$ the fluid viscosity and $W$ the out-of-plane width \cite{acheson1990elementary}. Here $e_{ij}$ is the symmetric part of the velocity gradient tensor (matrix), so
\begin{align}
    2 e_{ij}^2 &= 2 \partial_x u^2 + (\partial_y u + \partial_x v)^2 + 2\partial_y v^2.  
\end{align}
Plugging in $u$ and $v$ in terms of $\psi$ and nondimensionalizing the power by $\mu W\kappa^2/L_x^2$,
\begin{align} 
   \mbox{Power} 
   &= \int_0^1 \int_0^1\left(\partial_{xx}\psi^2 + \partial_{yy}\psi^2 -2 \partial_{xx}\psi\partial_{yy}\psi + 4\partial_{xy}\psi^2\right)  dx dy. \label{nondimPower}
\end{align}

\section{Numerical method}
\label{SectionIII}
\subsection{Non-uniform Mesh}
\label{SectionIIIA}
\begin{figure}
    \centering
    \includegraphics[width= 5 in]{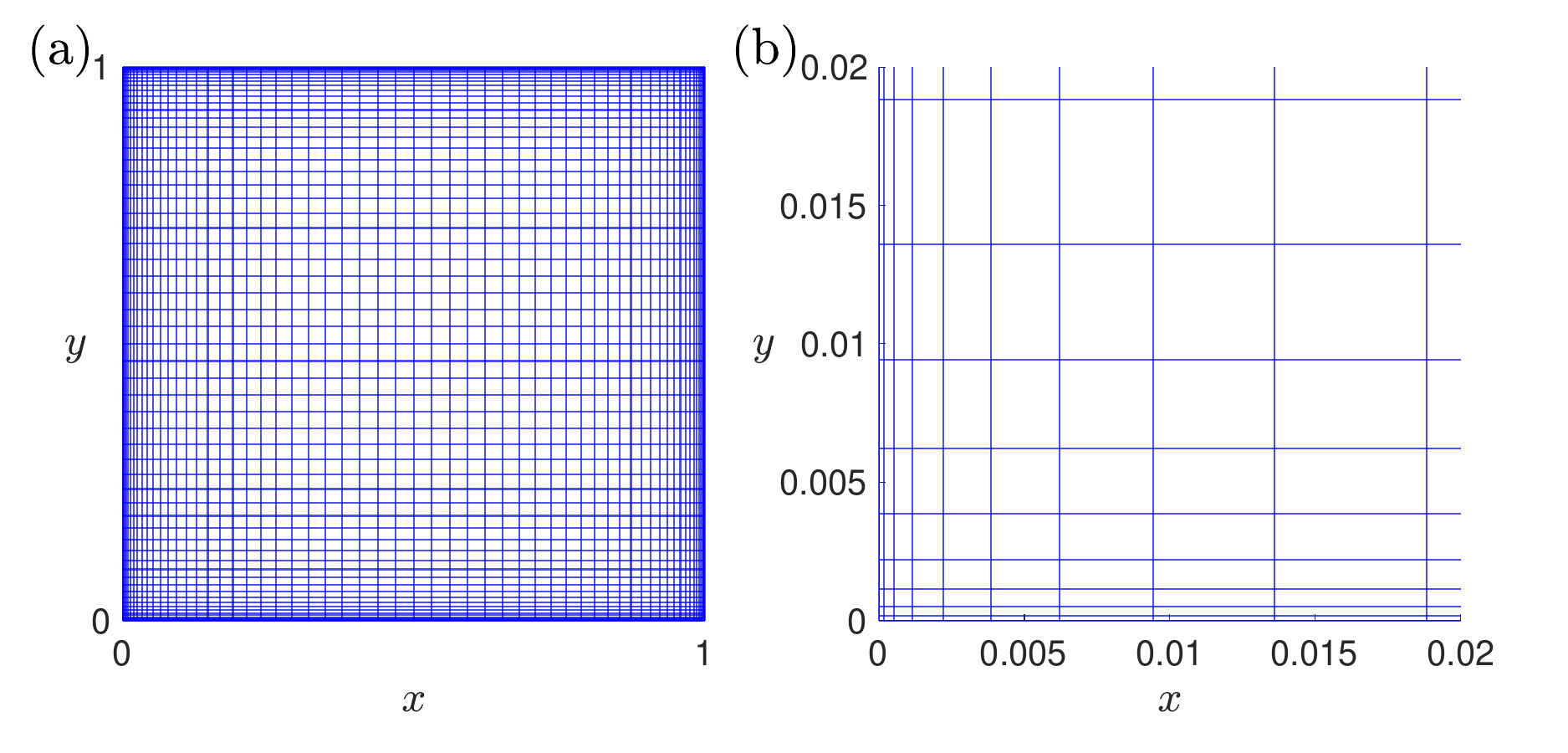}
    \caption{Example of the non-uniform computational mesh with $m, n = 64$. The figure on the right is a zoomed in view of left corner of the left figure, where we can see that as we get closer to the boundaries we have more grid lines.}
    \label{fig:NonUniformMesh}
\end{figure}

We study a channel with square aspect ratio because it provides a simple  baseline geometry. We employ a similar structured, non-uniform mesh technique  as used in \cite{alben2021collective} to efficiently capture the flow details near the top and bottom walls while constraining the overall number of unknowns and total computational cost. This approach provides finer resolution near the boundaries to capture thin boundary layers at higher $Pe$ values, while reducing grid density in regions requiring less detail. Figure \ref{fig:NonUniformMesh} shows the non-uniform grid spacing that is achieved by modifying a uniform grid with a sinusoidal term that introduces clustering near the boundaries: 
\begin{equation*}
    x_{\text{non-uniform}} = x_{\text{uniform}} - \lambda \cdot \frac{1}{2\pi}\cdot \sin(2\pi x_{\text{uniform}}), \hspace{0.3 in} 0 \leq x_{\text{uniform}} \leq L_{x}
\end{equation*}
\begin{equation*}
    y_{\text{non-uniform}} = y_{\text{uniform}} - \lambda \cdot \frac{1}{2\pi}\cdot \sin(2\pi y_{\text{uniform}}), \hspace{0.3 in} 0 \leq y_{\text{uniform}} \leq H
\end{equation*}
where $x_{\text{uniform}}$ and $y_{\text{uniform}}$ are vectors containing $m+1$ and $n+1$ evenly spaced points between $0$ and $1$ respectively. The sinusoidal term causes local stretching and compression in the grid, depending on the position along the coordinate axis. This clustering effect is controlled by the grid stretching factor $\lambda$, which we set to 0.997. Near the endpoints of the grid (i.e., at $ x_{\text{uniform}} = 0 $ and 1 and $ y_{\text{uniform}} = 0 $ and 1), $dx_{\text{non-uniform}}/dx_{\text{uniform}}$ and $dy_{\text{non-uniform}}/dy_{\text{uniform}} = 1-\lambda$ which is small, so grid spacings are decreased. At the center of the grid (i.e., at $x_{\text{uniform}} = 0.5 $ and $y_{\text{uniform}} = 0.5 $), the grid spacing is that of the uniform grid multiplied by $1+\lambda \approx 2$. Thus $\lambda$ controls how much the uniform grid is stretched or compressed. After testing convergence on several mesh sizes, we chose $m = n = 256$ for flows computed with $Pe = 2^{12}, 2^{13}, 2^{14}$, and $2^{15}$ and $m = n = 384$ for flows computed with $Pe = 2^{16}$ and $2^{17}$. We provide information about the effects of different mesh sizes in appendix \ref{AppendixA}. 

\subsection{Computing the gradient of $Nu$ using the adjoint method}
\label{SectionIIIB}

We compute the gradient of $Nu$ (\ref{eqn:Nu}) with respect to our design parameters, the entries of $\mathbf{B}$, using the adjoint method \cite{martins2021engineering}. First, we normalize $\psi$ to ensure the power (\ref{nondimPower}) is fixed across different flow configurations. This is done numerically by discretizing the integrals in (\ref{nondimPower}) using the trapezoidal rule and approximating derivatives with finite-difference matrices:
\begin{align}
   \mbox{Power} 
   &\approx \mathbf{\Psi}^T \mathbf{M} \mathbf{\Psi} \; ; \;
    \mathbf{M} \equiv 
    (\mathbf{Wt} \cdot \mathbf{D2x})^T \mathbf{D2x} +
   (\mathbf{Wt} \cdot \mathbf{D2y})^T \mathbf{D2y}
   -2
   (\mathbf{Wt} \cdot \mathbf{D2x})^T \mathbf{D2y} + 4
   (\mathbf{Wt} \cdot \mathbf{Dxy})^T \mathbf{Dxy}. \label{Mmat}
\end{align}
$\mathbf{Wt}$ is a matrix that contains weights corresponding to the trapezoidal rule, $\mathbf{D2x}$ and $\mathbf{D2y}$ are finite-difference matrices corresponding to second derivatives in $x$ and $y$ respectively, $\mathbf{Dxy}$ is the finite difference mixed partial derivative, and $\cdot$ indicates componentwise multiplication. To ensure the flow has (discretized) power $Pe^2$,
we write 
$\tilde{\mathbf{\Psi}} = \mathbf{\Psi}_1 + \mathbf{\Psi}_2$, and then
\begin{equation}
\mathbf{\Psi}_{\text{norm}}= \frac{\mbox{Pe}\, \tilde{\mathbf{\Psi}}}{\sqrt{(\tilde{\mathbf{\Psi}})^T\mathbf{M}\tilde{\mathbf{\Psi}}}} \label{Psi}
\end{equation}
has power $Pe^2$. We write the discretized advection diffusion equation (\ref{eqn:nondimadvecdiff}) in terms of a residual vector
\begin{equation}
    \mathbf{r}(\mathbf{\Psi}_{\text{norm}}(\mathbf{B}), \mathbf{T}) \equiv \mathbf{u} \cdot \mathbf{D1x} \mathbf{T} + \mathbf{v} \cdot \mathbf{D1y} \mathbf{T} - \mathbf{D2x} \mathbf{T}- \mathbf{D2y} \mathbf{T} = \mathbf{0}. \label{DiscRes}
\end{equation}
where
\begin{equation}
    \mathbf{u} = \mathbf{D1y} \mathbf{\Psi}_{\text{norm}} \quad ; \quad
    \mathbf{v} = -\mathbf{D1x} \mathbf{\Psi}_{\text{norm}}.
\end{equation}
Here $\mathbf{u}$,  $\mathbf{v}$, and $\mathbf{T}$ are the values of $u, v$, and $T$ on the numerical grid, arranged as vectors. $\mathbf{D1x}$ and $\mathbf{D1y}$ are the finite difference matrices for the first derivatives in the $x$ and $y$ directions respectively.
We consider (\ref{DiscRes}) as a system of implicit equations for $\mathbf{T}$ as a function of $\mathbf{B}$, the mode coefficients $\{B_{jk}\}$ that define the flow, arranged as a vector. Given the implicit dependence of the Nusselt number $Nu$ on the design variables $\mathbf{B}$, the total derivative of $Nu$ with respect to $\mathbf{B}$ can be expressed using the chain rule as:
\begin{equation}
\label{eqn:dNudB}
    \frac{dNu}{d \mathbf{B}} = \frac{d Nu}{d \mathbf{T}} \frac{d\mathbf{T}}{d \mathbf{B}} = \frac{dNu}{d\mathbf{T}} \frac{d \mathbf{T}}{d\mathbf{\Psi}_{\text{norm}}} \frac{d \mathbf{\Psi}_{\text{norm}}}{d \mathbf{B}}.
\end{equation}
Here $ \frac{dNu}{d \mathbf{T}} $ has dimensions $ 1 \times (m+1)(n+1) $, $ \frac{d \mathbf{T}}{d \mathbf{\Psi}_{\text{norm}}} $ is of size $ (m+1)(n+1) \times (m+1)(n+1) $, and $ \frac{d \mathbf{\Psi}_{\text{norm}}}{d \mathbf{B}} $ has dimensions $ (m+1)(n+1) \times M_2N_2 $. Consequently, the overall derivative $ \frac{dNu}{d \mathbf{B}} $ has dimensions $ 1 \times M_2N_2$.

To the find the total derivative $\frac{d \mathbf{T}}{d\mathbf{\Psi}_{\text{norm}}}$, we need to consider the residual equation (\ref{DiscRes}).
Assuming that we are at a point where $\mathbf{r}(\mathbf{\Psi}_{\text{norm}}(\mathbf{B}), \mathbf{T}) = 0$, any perturbation in $\mathbf{\Psi}_{\text{norm}}(\mathbf{B})$ must be accompanied by a perturbation in $\mathbf{T}$ such that the advection-diffusion equation remains satisfied. Thus, the differential of the residuals can be written as 
\begin{equation} \label{eqn:dr}
    d\mathbf{r} = \frac{d \mathbf{r}}{d \mathbf{\Psi}_{\text{norm}}} \triangle \mathbf{\Psi}_{\text{norm}} + \frac{d \mathbf{r}}{d\mathbf{ T}} \triangle \mathbf{T} = 0
\end{equation}
The total derivative $\frac{dNu}{d\mathbf{B}}$ (\ref{eqn:dNudB}) represents the effect that a perturbation on $\mathbf{B}$ has on $Nu$ subject to the constraint of satisfying the advection-diffusion equation, which can be achieved with the appropriate variation in $T$ \cite{martins2021engineering}. We rearrange (\ref{eqn:dr}) to get
\begin{equation} \label{eqn:drdTdTdPsi}
    \frac{d \mathbf{r}}{d \mathbf{T}}\frac{d \mathbf{T}}{d \mathbf{\Psi}_{\text{norm}}} = - \frac{d \mathbf{r}}{d \mathbf{\Psi}_{\text{norm}}}
\end{equation}
Solving (\ref{eqn:drdTdTdPsi}) for $\frac{d \mathbf{T}}{d \mathbf{\Psi}_{\text{norm}}}$, and substituting into (\ref{eqn:dNudB}) we get
\begin{equation}
   \frac{d\mbox{Nu}}{d\mathbf{B}} = -\frac{d\mbox{Nu}}{d\mathbf{T}}\frac{d\mathbf{r}}{d\mathbf{T}}^{-1}\frac{d\mathbf{r}}{d\mathbf{\Psi}_{\text{norm}}}\frac{d\mathbf{\Psi}_{\text{norm}}}{d\mathbf{B}} \label{eqn:dNudB1}
\end{equation}
The adjoint method changes the linear system that is solved to compute the total derivatives. Looking at (\ref{eqn:dNudB1}), we see that instead of solving a linear system with $\frac{d \mathbf{r}}{d \mathbf{\Psi}_{\text{norm}}}\frac{d  \mathbf{\Psi}_{\text{norm}}}{d \mathbf{B}}$ on the right-hand side, we can solve it with $\frac{d\text{Nu}}{d\mathbf{T}}$ on the right-hand side. The adjoint method reduces the computational cost because it allows us to compute $\frac{d\text{Nu}}{d\mathbf{B}}$ without explicitly computing or inverting the Jacobian matrix $\frac{d \mathbf{T}}{d \mathbf{\Psi}_{\text{norm}}}$, which is a dense $(m+1)(n+1) \times (m+1)(n+1)$ matrix. The adjoint method reduces computational cost by converting a problem that scales with the number of design variables into a problem that scales with the number of output functions, in this case, just $\mathrm{Nu}$. This makes the adjoint method especially efficient when the number of design variables is much larger than the number of outputs.
This corresponds to computing the product of the first two terms on the right-hand side,
\begin{equation} \label{eqn:etaT}
    \eta^{T} = \frac{d\mbox{Nu}}{d\mathbf{T}}\frac{d\mathbf{r}}{d\mathbf{T}}^{-1}
\end{equation}
where $\eta$ is the adjoint vector. 
Multiplying both sides of (\ref{eqn:etaT}) by $\frac{d \mathbf{r}}{d \mathbf{T}}$ on the right and taking the transpose of both sides, we obtain the adjoint equation \cite{martins2021engineering},
\begin{equation}
    \frac{d \mathbf{r}}{d \mathbf{T}}^{T} \eta = \frac{d Nu}{d \mathbf{T}}^{T} \label{eqn:adjoint}
\end{equation}
The solution $\eta$ is then used to compute the total derivative of Nu with respect to $\mathbf{B}$: 
\begin{equation}
    \frac{d\mbox{Nu}}{d\mathbf{B}} =-\mathbf{\eta}^T\frac{d\mathbf{r}}{d\mathbf{\Psi}_{\text{norm}}}\frac{d\mathbf{\Psi}_{\text{norm}}}{d\mathbf{B}}
\end{equation}

\subsection{BFGS optimization algorithm}
\label{SectionIIIC}
After computing $\frac{d \text{Nu}}{d \mathbf{B}}$ using the adjoint method, we employ the Broyden-Fletcher-Goldfarb-Shanno (BFGS) algorithm to find optimal flows. BFGS is useful for this optimization because it uses second-order information (through an approximation of the Hessian matrix) to improve optimization efficiency, without the computational cost of directly inverting the full Hessian matrix \cite{nocedal2006numerical}. This makes it faster and more memory efficient than methods that require the explicit computation of second-order derivatives, such as Newton's method \cite{broyden1970convergence}. We initialize the approximate inverted Hessian, $\mathbf{H}_{0}^{*}$, as the identity matrix. Another approach would be to compute the true inverted Hessian at the initial point. This would start the algorithm off with more efficient steps, but comes at an initial cost of computing the true Hessian and inverting it. 
We compute the current search direction, $p_{k}$, by using the current approximation $\mathbf{H}_{k}^{*}$ of the inverted Hessian matrix: 
\begin{equation}
    p_{k} = - \mathbf{H}_{k}^{*} \cdot \Big(\frac{d Nu}{ d\mathbf{B}}\Big)_{k}
\end{equation}
where $\Big(\frac{dNu}{d\mathbf{B}}\Big)_{k}$ is the gradient of $Nu$ at the current point $\mathbf{B}_{k}$. Then we perform a line search to determine an optimal step size $\alpha_{k}$. The line search ensures that the new point $\mathbf{B}_{k+1}$ reduces the objective function sufficiently (and ideally, improves the convergence rate). The line search typically involves an inexact search over the interval and can be performed using a variety of techniques such as backtracking or the Wolfe conditions \cite{fletcher1987practical}. In our case, we use an exponential backtracking line search. We update the step size as
\begin{equation}
    \alpha_{j} = 2^{-j} 
\end{equation}
in an inner loop iterated by $j$. $j$ starts at $-3$ and increases up to a maximum of $14$. At each iteration, we calculate $Nu_{j+1}$ using: 
\begin{equation}
    \mathbf{B}_{j+1} = \mathbf{B}_{j} + \alpha_{j} p_{k}
\end{equation}
If $Nu_{j+1} > Nu_{j}$, then we accept $\mathbf{B}_{j+1}$, and set $\mathbf{B}_{k+1} = \mathbf{B}_{j+1}$. If not, we increment $j$, halving the step size. We continue the line search until we find an improvement on $Nu$, or else we accept the last (smallest) step. Having chosen the step size, we update the
inverted Hessian approximation using the BFGS update formula \cite{martins2021engineering}.
 We continue the iteration until the norm of the gradient $\Big|\Big| \Big(\frac{d Nu}{ d\mathbf{B}}\Big)_{k}\Big|\Big|$ is smaller than a chosen tolerance, and accept the computed velocity field as an optimal flow.  

 We initialize the BFGS optimization using a low-dimensional representation (fewer modes) and progressively increase the resolution by adding more modes throughout the optimization process. To maintain continuity, the added mode coefficients start at zero and change as the iteration proceeds. We increase the number of modes by one after a fixed number of iterations until reaching a predefined number of ending modes. For the flows generated on the $256 \times 256$ mesh, we used 6 random seeds, 16, 24, and 28 starting modes, and 32 ending modes. For the flows computed on the $384 \times 384$ mesh, we used 2 random seeds, 24, 36, and 42 starting modes, and 48 and 60 ending modes. For both meshes, we increased the number of modes by 1 after 2, 5, 10, or 20 iterations. This incremental mode update strategy proved to have several benefits. For one, the low-mode representations capture the dominant, large-scale structures of the flows. These initial, smoother approximations (which are composed of low-order Chebyshev polynomials) lead to a simplified, lower-dimensional optimization problem that is less likely to get trapped in poor local minima. Once an approximate solution is found in this simplified space, higher modes can refine it without starting from scratch. Additionally, we found that using fewer modes in the beginning reduced both memory usage and computational cost per optimization step at the outset. Gradually increasing the number of modes allowed the optimization algorithm to refine the solution while increasing detail only where needed. We usually saw convergence to better flows by incrementally increasing the number of modes than by using the full number of modes throughout the iteration. We list the numbers of starting and ending modes and update schedules we used in appendix \ref{AppendixC}.

The main computational cost is solving the advection-diffusion equation using MATLAB's backslash operator, which is done once for each iterate $j$ of the inner loop. Once the new step is accepted, the adjoint variable is computed in order to compute the gradient, with a cost of one solve of the adjoint of the advection-diffusion matrix (\ref{eqn:adjoint}). The costs of the other steps of the BFGS algorithm---updating the inverse Hessian approximation and computing the new search direction---are relatively insignificant.
In our study, the optimizer ran for 3,000–7,000 iterations, with the gradient norm decreasing by three orders of magnitude, and each run took between 24 and 72 hours on a single CPU.

\section{Results}
\label{SectionIV}
\subsection{Nearly-unidirectional optimal  flows}
\label{SectionIVA}
\begin{figure}
    \centering
    \includegraphics[scale = 0.5]{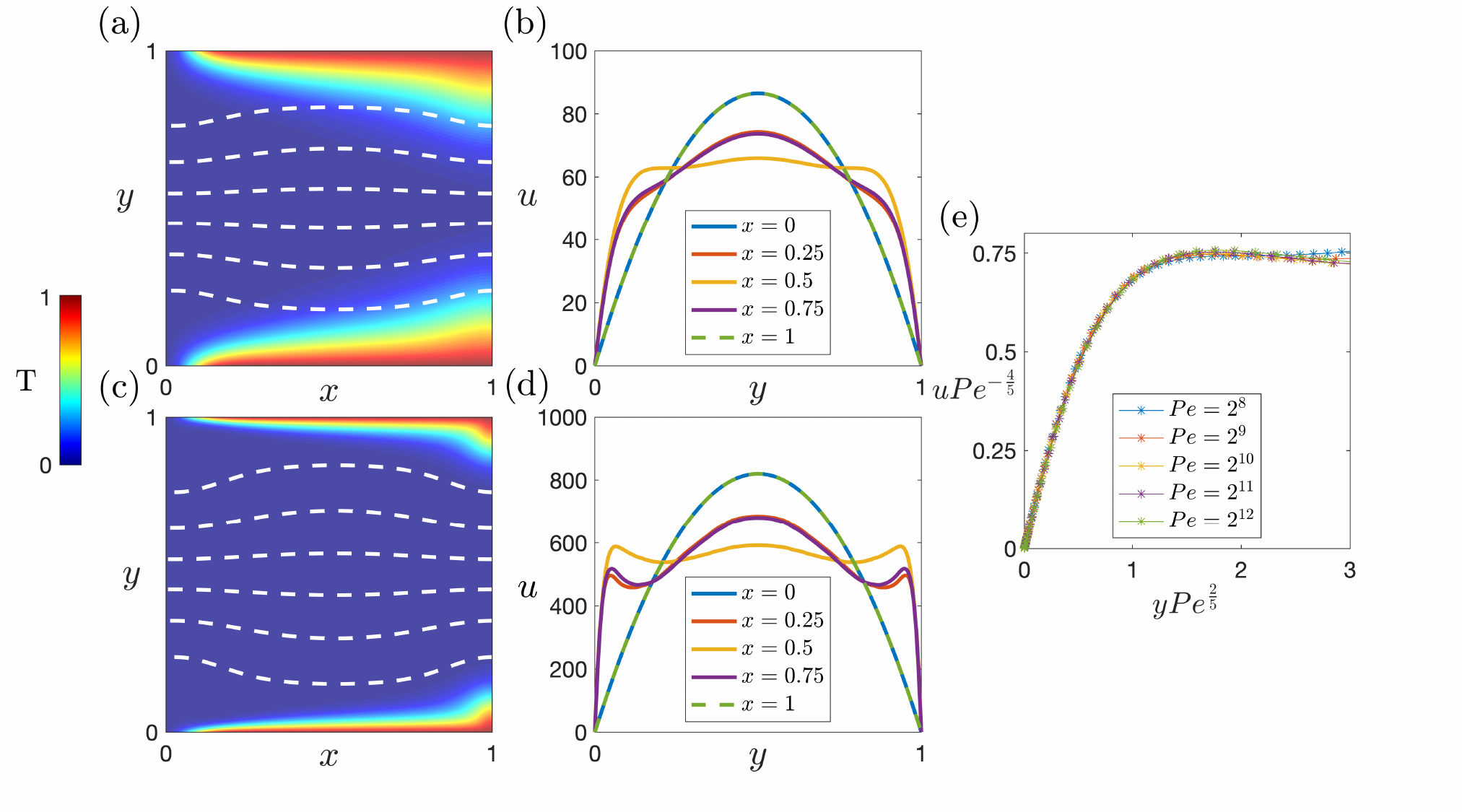}
    \caption{Panels (a) and (c) show contour plots of the temperature solution with the white, dashed lines showing the streamlines of $\Psi_{norm}$ for the optimal uniform flow at $Pe = 2^{8}$ and $Pe = 2^{12}$. Panels (b) and (d) show the horizontal velocity profile at $x = [0, 0.25, 0.5, 0.75, 1]$. Panel (e) shows the re-scaled horizontal velocity profile of the optimal uniform flows for $Pe$ ranging from $2^{8}$ to $2^{12}$ near the lower wall at $x = \frac{1}{2}$.}
    \label{fig:OptUnifFlows}
\end{figure}

In \cite{Alben2017Channel}, we computed optimal flows for the same physical model by solving the Euler-Lagrange equations, a coupled system of fourth-order nonlinear PDEs derived by the calculus of variations. We used a uniform mesh and found optimal flows for a range of Péclet values. Between $Pe = 2^{8} - 2^{11}$, the optimal flows were approximately unidirectional
except very close to the inflow and outflow boundaries. Specifically, we found $(u(x,y),v(x,y)) \approx (u(y), 0)$ for $0.2 \lessapprox x \lessapprox 0.8$. In figure~\ref{fig:OptUnifFlows} we show our computed optima with the present adjoint method, which agree well with the previous results. We find that for $Pe < 2^{12}$, the best computed optima (those with the largest Nu) are approximately unidirectional flows and share the same structure dominated by nearly uniform horizontal flow through the channel. Figure~\ref{fig:OptUnifFlows} illustrates this agreement by showing representative optimal unidirectional flows we found at $Pe = 2^{8}$ and $Pe = 2^{12}$ that were computed using a $128 \times 128$ mesh. 

In panel (a) we show a contour plot of the temperature field for the best performing unidirectional flow for $Pe = 2^{8}$, where the overlaid white dashed lines show the streamlines of the flow. This flow has a Nusselt number of 12.802 for $\psi_{\text{top,min}} = 0.05$. The optimization was
initialized with $8$ modes and increased by $1$ every $10$ iterations until reaching $16$ modes. Panel (b) shows the horizontal velocity profiles of the flow at $x = 0, 0.25, 0.5, 0.75, 1$. We see that for $x = 0$ and $x = 1$, the profiles are overlapping and parabolic representing the Poiseuille flow at the inlet and outlet of the channel. At values closer to the center of the channel like $x = 0.25, 0.5,$ and $0.75$, we observe that the flow has a flatter profile which is almost constant away from the boundary layers. We notice that the profile is most uniform at the center of the channel, when $x = 0.5$. In panel (c), we see the contour plot of the temperature field for a local optimum at $Pe = 2^{12}$, where the white lines again show the streamlines of the flow. This flow at achieves a Nusselt value of $37.395$ with $\psi_{\text{top,min}}$ = 0.05. This flow was found by starting the optimization routine with $4$ modes and increasing by $1$ every $2$ iterations until reaching $16$ modes. Interestingly, we observed that across $Pe = 2^{8} - 2^{11}$, the best flows are approximately unidirectional with boundary layers near the walls, similar to those shown in figure~\ref{fig:OptUnifFlows}. Panel (e) shows that dividing $u$ by $Pe^{4/5}$ and $y$ by $Pe^{-2/5}$ collapses the horizontal velocity profiles in the boundary layers at $x = 0.5$ into a single curve, in agreement with \cite{Alben2017Channel} where the power law exponents are derived theoretically. 

At $Pe = 2^{12}$ we find a remarkable change in the optimal flows that was not found by \cite{Alben2017Channel}. In that study, a uniform mesh was used to compute the 2D flow optima. The mesh resolution was constrained by the ill-conditioning of the fourth-derivative finite difference matrices. With only second-derivative matrices in this study, we can use a much finer mesh near the walls and resolve the optimal flows for $2^{12} < Pe \leq 2^{17}$.
Above a critical $Pe$ $\approx 2^{12}$, the optimal flows exhibit wavy streamline patterns with internal eddies and strong vertical heat transport. Above the critical $Pe$ the wavy flows outperform their unidirectional counterparts in heat transferred for a given power budget. This bifurcation in the type of optimal flow for heat transfer in a channel is the main finding of this work.  

\subsection{Wavy optimal flows}
\label{SectionIVB}
\subsubsection{Optimal flows}
\label{SectionIVB1}
\begin{figure}
\hspace{-0.1 in}\includegraphics[width=1 \linewidth]{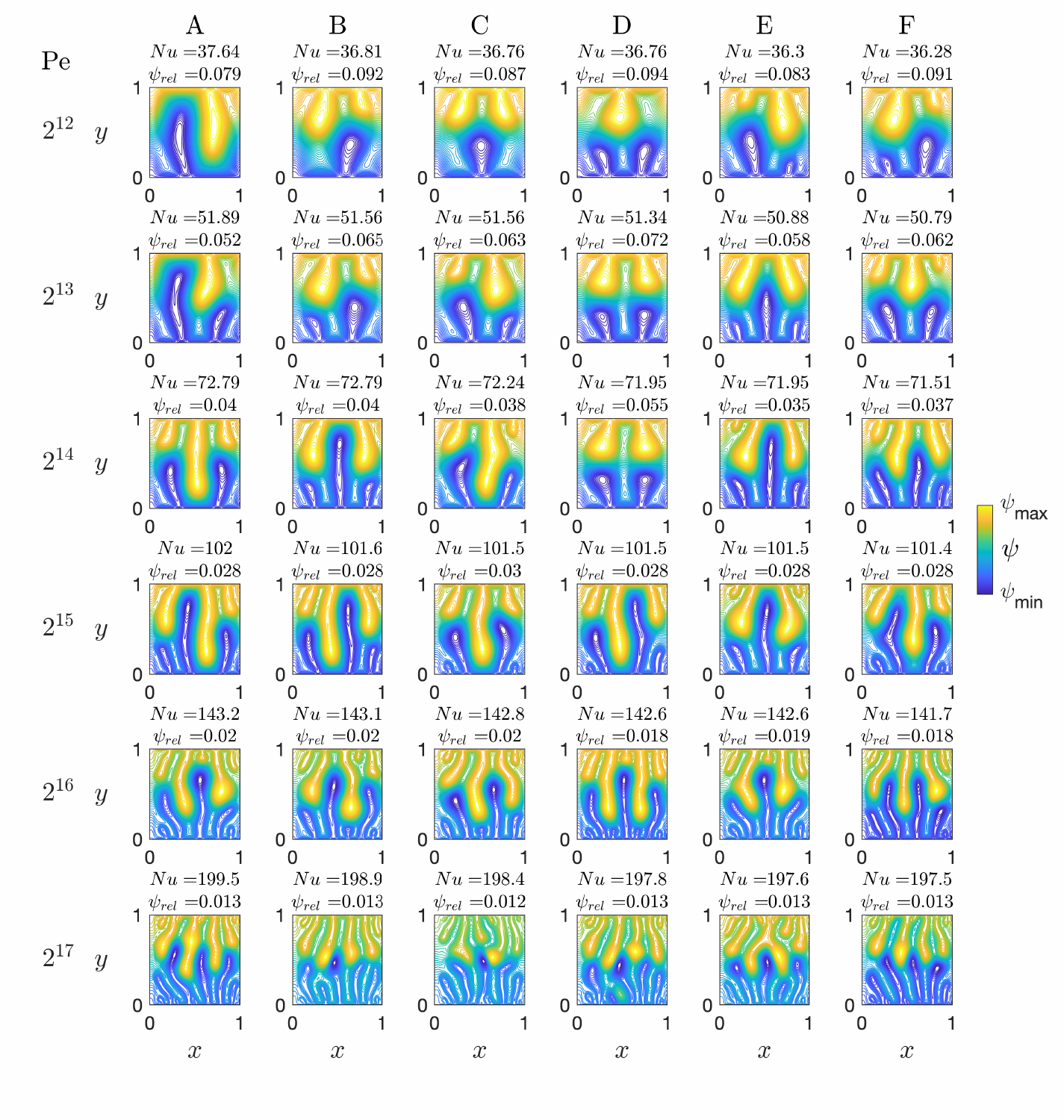}
\vspace{-0.6 in}
   \caption{Contour plots of the stream functions for the different optima found for $Pe = 2^{12}$--$2^{17}$ (in rows 1--6 respectively). We use a $256 \times 256$ mesh for $Pe = 2^{12} - 2^{15}$ and a $384 \times 384$ mesh for $Pe = 2^{16} - 2^{17}$. }  \label{fig:OptimalWavyFlows}
\end{figure}

The main novelty of this study is the wavy optimal flows that are observed from $Pe = 2^{12}$ to $2^{17}$. Streamlines for a selection of these flows are shown in figure~\ref{fig:OptimalWavyFlows}. For $Pe\geq 2^{12}$, the optimal flows differ noticeably from the approximately unidirectional ones at smaller $Pe$ shown in the last section. Figure~\ref{fig:OptimalWavyFlows} illustrates six of the top flows at each Pe. Column A shows the flows with the highest Nu, while columns B--F show other high-performing flows that were chosen to demonstrate a range of different flow patterns. The flows in columns B--F underperform the best flow by only 0--3$\%$, so this selection highlights a representative sample of the most effective flow configurations at each $Pe$ value. Across the entire set of optimal flows, we observe a consistent and striking pattern: long finger-like protrusions that extend deep into the channel from the top and bottom walls. The fingers become thinner and more numerous as $Pe$ increases.

We hypothesize that these convective fingers enhance heat transfer by providing a larger surface area for thermal transfer from hot to cold fluid in the channel. The flows take winding paths around the fingers as they travel through the channel. Along these winding paths the fluid undergoes greater thermal exchange, which corresponds to better heat transport at these $Pe$ compared to the unidirectional flows.
We observe flows that are approximately symmetric variants of each other, by 180-degree rotations or horizontal or vertical reflections, with approximately the same $Nu$. Examples are seen in columns C and D at $Pe$ = 2$^{12}$, B and C at $2^{13}$, and A and B at 2$^{14}$ and $2^{15}$. We will examine the effect of the structure and extent of the convective fingers later, in figure~\ref{fig:EffectofPTM}, when we show examples of flows with shorter, more localized fingers that remain close to the channel walls. These configurations are noticeably less effective at transferring heat, reinforcing the role of elongated fingers that reach well into the channel in driving more efficient heat transport. 

The emergence of finger-like structures in our optimal flow has some resemblance to previous studies on convection in horizontally periodic domains, in which the heat transfer is primarily vertical rather than horizontal. One set of studies considered double-diffusive convection and fingering instabilities. In particular, Yang et al. (2016) demonstrated that in double-diffusive systems, convection can transition from large-scale rolls to vertically oriented salt fingers when the ratio of temperature to salinity buoyancy force is sufficiently large \cite{yang2016convection}. These convective fingers, arising in natural rather than forced convection and with different physical mechanisms and boundary conditions, correspond to a modest increase in the rate of salinity transfer in the fluid. The scaling analyses and numerical results from Yang et~al. show that finger-dominated regimes in double-diffusive systems boost vertical salinity transport \cite{yang2016scaling}. 

Vertically-elongated convective structures were also observed in the optimization study \cite{alben2023transition} which used an unconstrained optimization framework similar to the present study, but in a horizontally periodic domain with vertical heat transfer as in several previous studies of wall-to-wall heat transfer \cite{hassanzadeh2014wall,tobasco2017optimal,motoki2018maximal,souza2020wall} and the double-diffusive convection work mentioned above. \cite{alben2023transition} found that as $Pe$ increases, optimal flows undergo a transition from simple convective rolls to elongated, branched, asymmetric, and heterogeneous structures. These branching flows corresponded to an increased $Nu$ power law scaling exponent and decreased peak velocity scaling exponents. Our results resemble this progression: as $Pe$ increases, the optimal flow fields develop more numerous and thinner fingers. Such structures appear to enhance convective transport without excessive velocity gradients and viscous dissipation.

Above each flow in figure \ref{fig:OptimalWavyFlows} we show $\psi_{\text{rel}}$, the normalized flux through the channel, defined as
\begin{equation}
    \psi_{\text{rel}} = \frac{\psi_{\text{top,min}} \cdot \sqrt{12}}{Pe}
\end{equation}
This dimensionless quantity is the flux through the channel ($\psi_{\text{top,min}}$) relative to the maximum possible flux of an incompressible flow at this Pe, which is $Pe/\sqrt{12}$ and is attained by Poiseuille flow with power $Pe^2$, as shown in \cite{Alben2017Channel}. At $\psi_{\text{rel}} = 1$, we have Poiseuille flow (only $\psi_1$), while at $0 < \psi_{\text{rel}} < 1$ we have a combination of Poiseuille flow with the additional flow structures in $\psi_{2}$, such as eddies, vortices, recirculation regions, and convective fingers. We observe that the optimal solutions consistently favor lower values of $\psi_{rel}$, i.e. a strong $\psi_{2}$ relative to $\psi_{1}$. 
All of the best-performing flows have a $\psi_{rel} < 0.1$, with the optimal $\psi_{rel}$ value for each Péclet value decreasing monotonically as $Pe$ doubles. The optimal $\psi_{rel}$ value drops by $\sim 84\%$ as $Pe$ increases from $2^{12}$ to $2^{17}$. Flows with lower $\psi_{rel}$ values can have more vertical transport than the unidirectional flows, but less net fluid flux out of the channel. 
For the unidirectional optimal flows in figure \ref{fig:OptUnifFlows}, $\psi_{\text{rel}} = 0.70$ for $Pe = 2^9$, $\psi_{\text{rel}} = 0.62$ for $Pe = 2^{10}$, and $\psi_{\text{rel}} = 0.53$ for $Pe = 2^{11}$. Later, in figure~\ref{fig:CombinedComparisonPlots}b, we report power-law scaling behaviors of $\psi_{\text{rel}}$ with $Pe$ in both the unidirectional and wavy flow regimes.
The wavy flows have a longer interface between the hot and cold fluid, whereas the unidirectional flows have a greater flux through the channel, as shown by their significantly higher $\psi_{rel}$ values.

We also looked at how varying $\lambda$, the stretching factor, affected the results. As discussed in appendix \ref{AppendixA}, this revealed a trade-off between mesh resolution and matrix conditioning. Finer meshes offer higher spatial accuracy but often lead to ill-conditioned matrices, which can impede convergence. Conversely, coarser meshes result in better-conditioned systems and, in a few cases, yield slightly improved optimal flows, albeit at the cost of lower spatial accuracy. Figure \ref{fig:OptFlows_Lambda0.97} in appendix \ref{AppendixD} illustrates the most notable of these cases, where the resulting flow fields differ from those shown in figure \ref{fig:OptimalWavyFlows} and demonstrate moderately improved convergence. Given that $\lambda = 0.997$ generally gave nearly equal or larger $Nu$ than $\lambda = 0.97$, and similar flow patterns, we use $\lambda = 0.997$ for our results here.

\subsubsection{Optimal temperature solutions}
\label{SectionIVB2}
\vspace{-0.13 in}
\begin{figure}
\hspace{-0.1 in}\includegraphics[width=1 \linewidth]{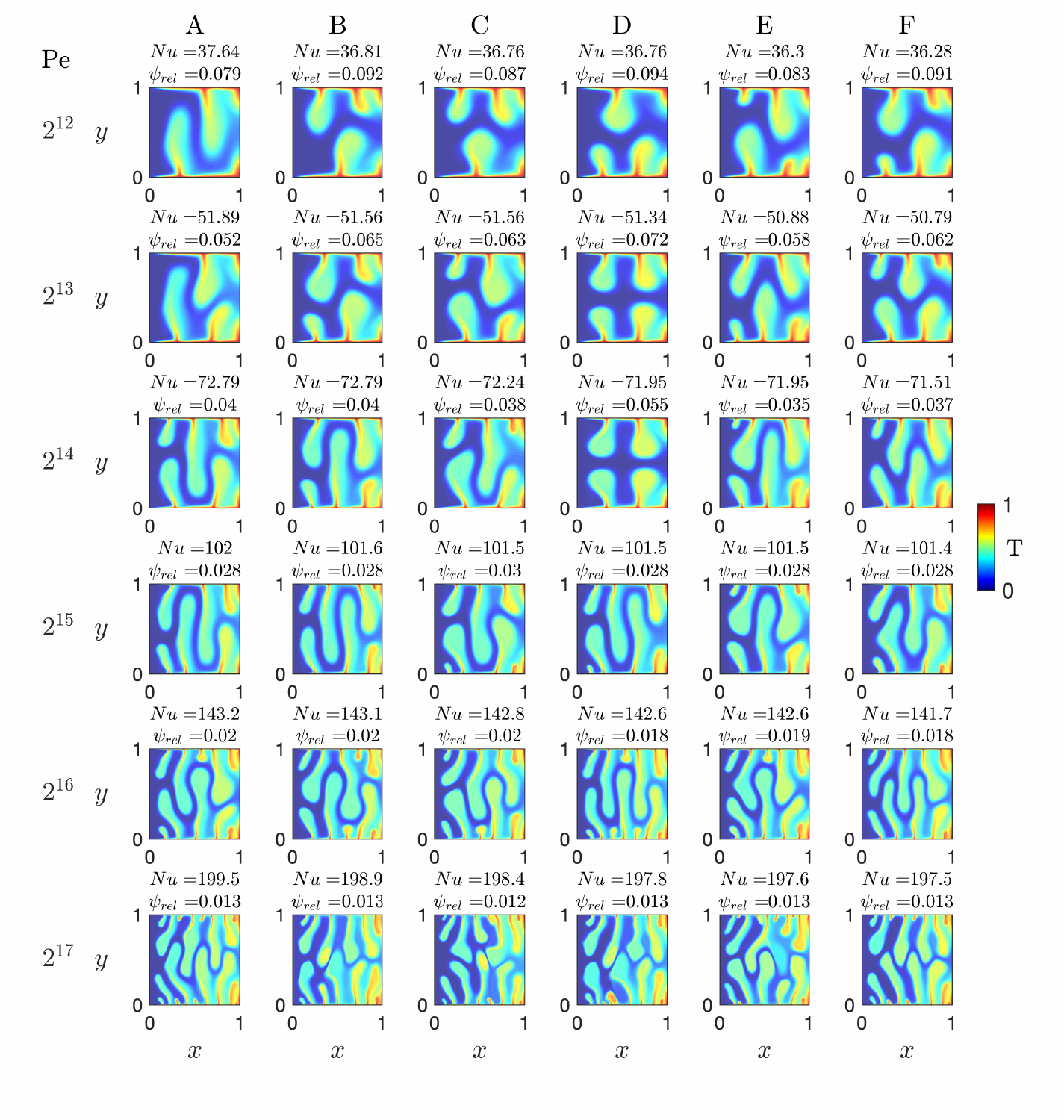}
\vspace{-0.6 in}
    \caption{\footnotesize Temperature fields for the flows in figure \ref{fig:OptimalWavyFlows}. }
    \label{fig:OptimalWavyFlowTemps}
\end{figure}

To illustrate the impact of the wavy optimal flow structures, in figure~\ref{fig:OptimalWavyFlowTemps} we present the temperature fields corresponding to the optimal flows in figure~\ref{fig:OptimalWavyFlows}. Corresponding to the convective fingers in the velocity field, we observe thin, elongated regions of hot fluid extending vertically from the top and bottom channel walls. These regions move hot fluid from the walls into the center of the channel. Notably, the cores of these thermal fingers appear relatively homogeneous in temperature, with values close to 0.5, indicating an effective mixing of hot and cold fluid within these regions. These thermal fields somewhat resemble those in natural convection,
but with more thermal homogenization \cite{Kellner_Tilgner_2014, yang2016scaling, yang2016convection}. 

\subsubsection{Flow Speed Visualization}
\label{SectionIVB3}
\begin{figure}
\hspace{-0.1 in}\includegraphics[width=1 \linewidth]{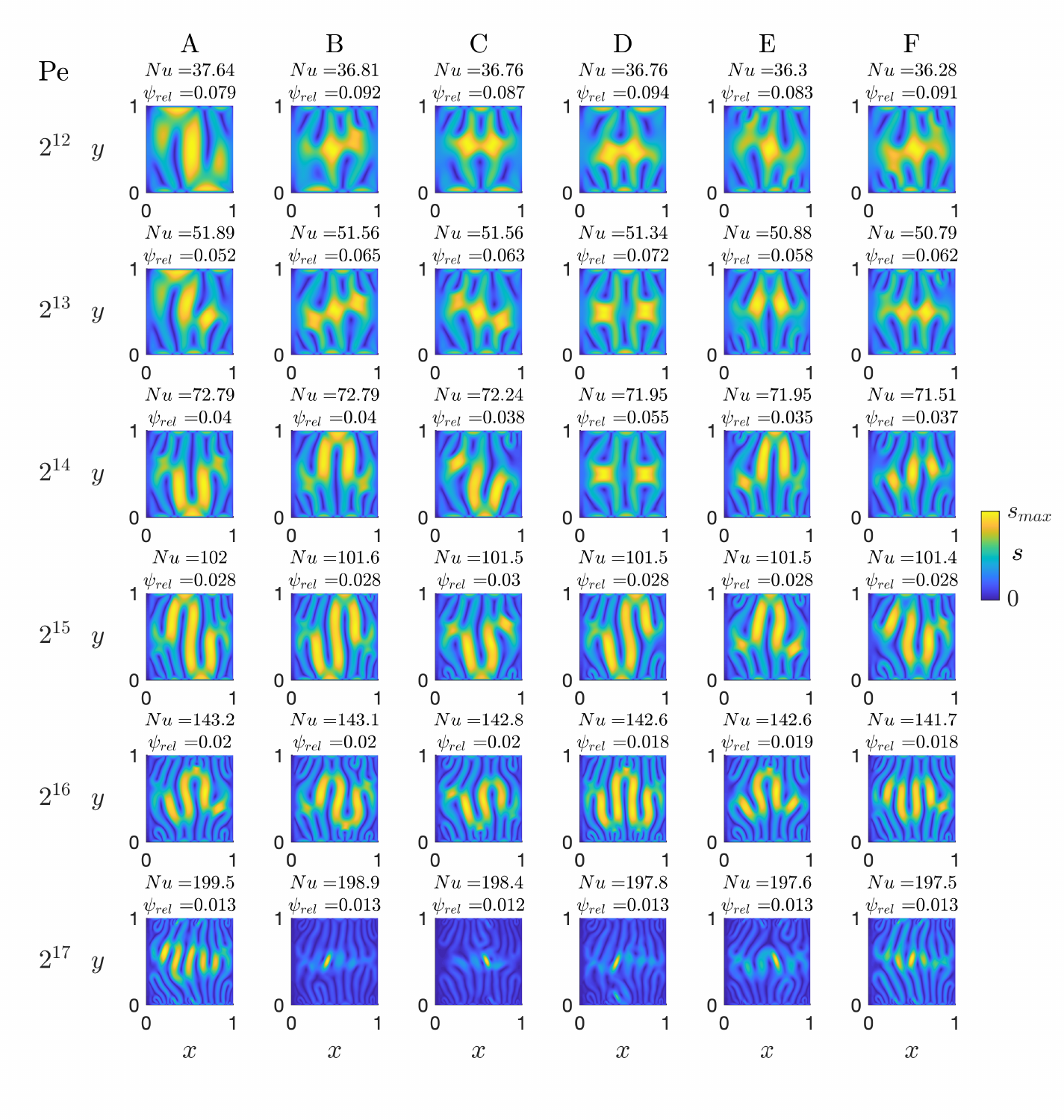}
\vspace{-0.6 in}
    \caption{\footnotesize Color plots of the flow speed $s$ for the flows in figure \ref{fig:OptimalWavyFlows}. 
    }
    \label{fig:OptimalWavyFlowSpeed}
\end{figure}

To better understand the optimized flow structures, we next examine the distribution of flow speed across the domain. In figure~\ref{fig:OptimalWavyFlowSpeed}, the maximum flow speeds generally occur near the center of the channel, where the flow navigates around the fingers closest to this region. In some cases there are also regions of high-speed flow close to the walls. For $Pe$ in the range $2^{12}$ to $2^{13}$, the highest velocities are concentrated near the channel center, approximately at $(x \approx 0.5, y \approx 0.5)$. Starting at $Pe = 2^{14}$, the flow behavior changes: as the fingers become longer and the flow winds more intricately around them, the regions of maximum velocity shift vertically along the centerline, appearing near $(x \approx 0.5, y \in [0.5, 1])$ or $(x \approx 0.5, y \in [0, 0.5])$, depending on the direction of finger extension. At $Pe = 2^{16}$, the fastest flow becomes more localized along the paths surrounding the central fingers. By $Pe = 2^{17}$, a range of distinct patterns emerge: some configurations exhibit multiple high-speed regions, others show highly localized zones of maximum velocity, and still others display consistently elevated speeds throughout the paths around the fingers. The regions of lowest speed tend to roughly trace the geometry of the fingers, indicating that the fingers themselves are composed of slow-moving fluid. 

\subsubsection{Effect of damping the interior flow near the upstream boundary}
\label{SectionIVB4}
\begin{figure}
\centering
\includegraphics[width=0.8 \linewidth]{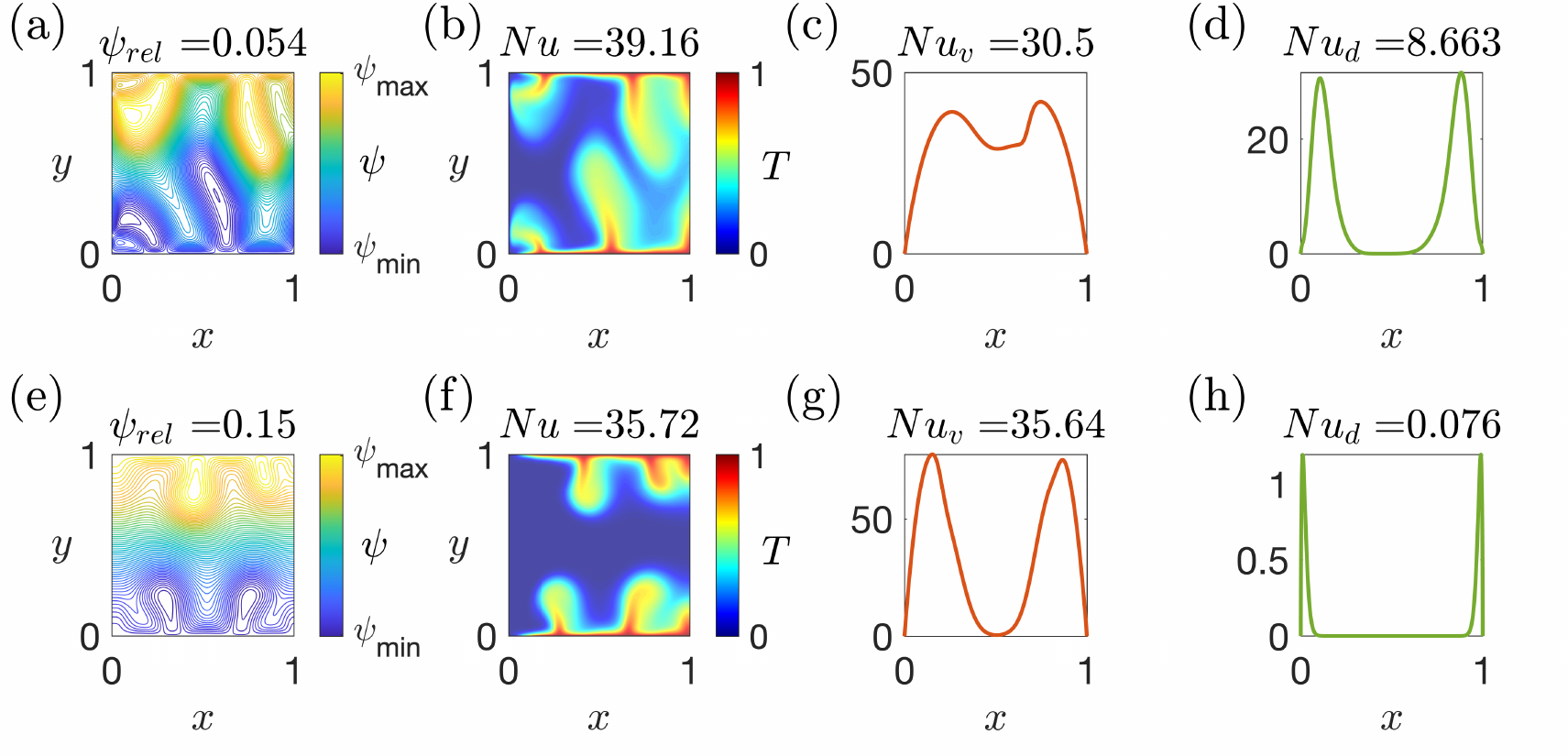}
    \caption{
    Optimal flows and temperature fields without the factor $(1 - e^{-\frac{x^4}{\delta^4}})$ applied to $\psi_2$ (panels (a)--(d)) and with the factor (panels (e)–(h)). 
    Contour plots of $\psi$ are shown in panels (a) and (e), temperature in (b) and (f), convective heat flux in (c) and (g), and conductive heat flux in (d) and (h), all for $Pe = 2^{12}$ and $\psi_{\text{top,min}} = 0.2$.}
    \label{EffectofSmallFlux}
\end{figure}

Figure \ref{EffectofSmallFlux} illustrates the effect of applying the factor $\left(1 - e^{-\frac{x^4}{\delta^4}}\right)$ to $\psi_2$. As discussed in section \ref{SectionIIB}, the purpose of this exponential factor is to smoothly suppress interior flow near the upstream boundary to ensure significant inward flow at the inlet and negligible heat conduction, thereby justifying the assumption that $T = 0$ at the inlet.  
We show two cases, where the top row (panels (a) -- (d)) corresponds to an optimal flow calculated without the factor, i.e.~with $\psi_{2}(x,y) = \Big[\sum_{j = 1}^{M_2}\sum_{k = 1}^{N_2} B_{jk} P_j(x)Q_k(y) \Big]$, while the bottom row (panels (e) -- (h)) corresponds to an optimal flow found with the same optimization parameters and with the factor $(1 - e^{-\frac{x^4}{\delta^4}})$ included in $\psi_{2}$. Panels (a) and (e) show the stream function contours for the optimal flows found with and without this factor, respectively. In panel (a), we see eddies close to the left boundary where cold fluid enters the channel.
These eddies exchange hot fluid near the top and bottom walls with cold fluid at the left boundary. Since there is little inward flow on this portion of the left boundary, it is more like a solid wall with temperature fixed at 0 by an external cooling source rather than an inlet. Without such additional cooling, these boundary eddies would raise $T$ above 0 there. Although these eddies are effective for heat exchange, they violate our assumption that $T \equiv$  0 at the left boundary due to the strong inward flow of cold fluid there.
In contrast, panel (e), where the factor $\left(1 - e^{-\frac{x^4}{\delta^4}}\right)$ is applied to $\psi_{2}$, shows streamlines that are more uniformly horizontal near the inlet. This modification ensures the presence of horizontal flow at the channel inlet, allowing solutions where heat conduction at the upstream boundary is minimal, and thus making it reasonable to assume that the temperature $T$ remains zero in that region.
Panels (b) and (f) show the temperature fields corresponding to panels (a) and (e) respectively. In panel (b), we see  recirculation zones caused by the eddies in panel (a) that draw hot fluid towards the left boundary, increasing the local temperature and producing sharp gradients in $T$ there. On the other hand, the temperature field in panel (f) shows that $T = 0$ at and near the left boundary. As the cold fluid moves downstream, it is progressively heated within the channel. In panels (c)--(d) and (g)--(h), we show the relative heat transfer due to convection ($Nu_{v})$ vs. conduction at the upstream boundary ($Nu_{d})$. We define the convective and diffusive heat transfer contributions as
\begin{equation}
    Nu_v = \int_{0}^{1} \left( T \cdot \partial_{y} \psi \,\big|_{x=1} - T \cdot \partial_{y} \psi \,\big|_{x=0} \right) \, dy \quad ; \quad Nu_{d} = \int_{0}^{1} (\partial_{x}T\big|_{x =0}) dy.
\end{equation}
We show that $Nu = Nu_{v} + Nu_{d}$ by integrating the steady advection-diffusion equation over the domain and applying the divergence theorem:  
\begin{equation*}
    \int_{\Omega} u \cdot \nabla T - \nabla^{2}T dA = 
    \int_{\partial \Omega} (T u\cdot \hat{n} - \partial_{n}T) ds = 0
\end{equation*}
Evaluating the boundary terms, we have: 
\begin{equation*}
    \underbrace{\int_{0}^{1} \left( T \cdot \partial_{y} \psi \,\big|_{x=1} - T \cdot \partial_{y} \psi \,\big|_{x=0} \right) \, dy}_{\text{$Nu_{v}$}} + \underbrace{\int_{0}^{1} (\partial_{y}T|_{y = 0} - \partial_{y}T|_{y= 1}) dx}_{\text{$-Nu$}} + \underbrace{\int_{0}^{1} (\partial_{x}T\big|_{x =0}) dy}_{\text{$Nu_{d}$}} = 0
\end{equation*}

In reality, heat transfer by conduction at the inlet should be negligible. Panels (c) and (d) present the convective heat flux and the conductive heat fluxes at the left boundary, respectively, and show that $78\%$ of the heat transfer occurs through convection and $22\%$ through conduction at the left boundary, corresponding to the eddies in panel (a). The corresponding flux profiles following the application of the factor $\left(1 - e^{-\frac{x^4}{\delta^4}}\right)$ to $\psi_2$ are shown in panels (g) and (h). After applying the factor, the conductive heat flux is reduced from 22\% to just 0.2\%. The peaks in conductive heat flux at the top and bottom of the left boundary decrease by a factor of approximately 30 from panel (d) to (h). Incidentally, the peaks in convective heat flux increase by a factor of about 2 from panel (c) to (g), because the flows in (a) and (e) are different, but this is irrelevant to the main point, which is that the conductive flux at the left boundary has been made insignificant. The factor $(1 - e^{-\frac{x^4}{\delta^4}})$ guides the optimization process towards flow configurations with inflow rather than eddies at the left boundary. The Nusselt values found after this modification are somewhat lower but the solutions obey our assumption that $T = 0$ at the inflow boundary without additional sources of cooling there.

\subsubsection{Effects of varying $\psi_{\text{top,min}}$}
\label{SectionIVB5}
\begin{figure}
\centering\includegraphics[width= 0.9 \linewidth]{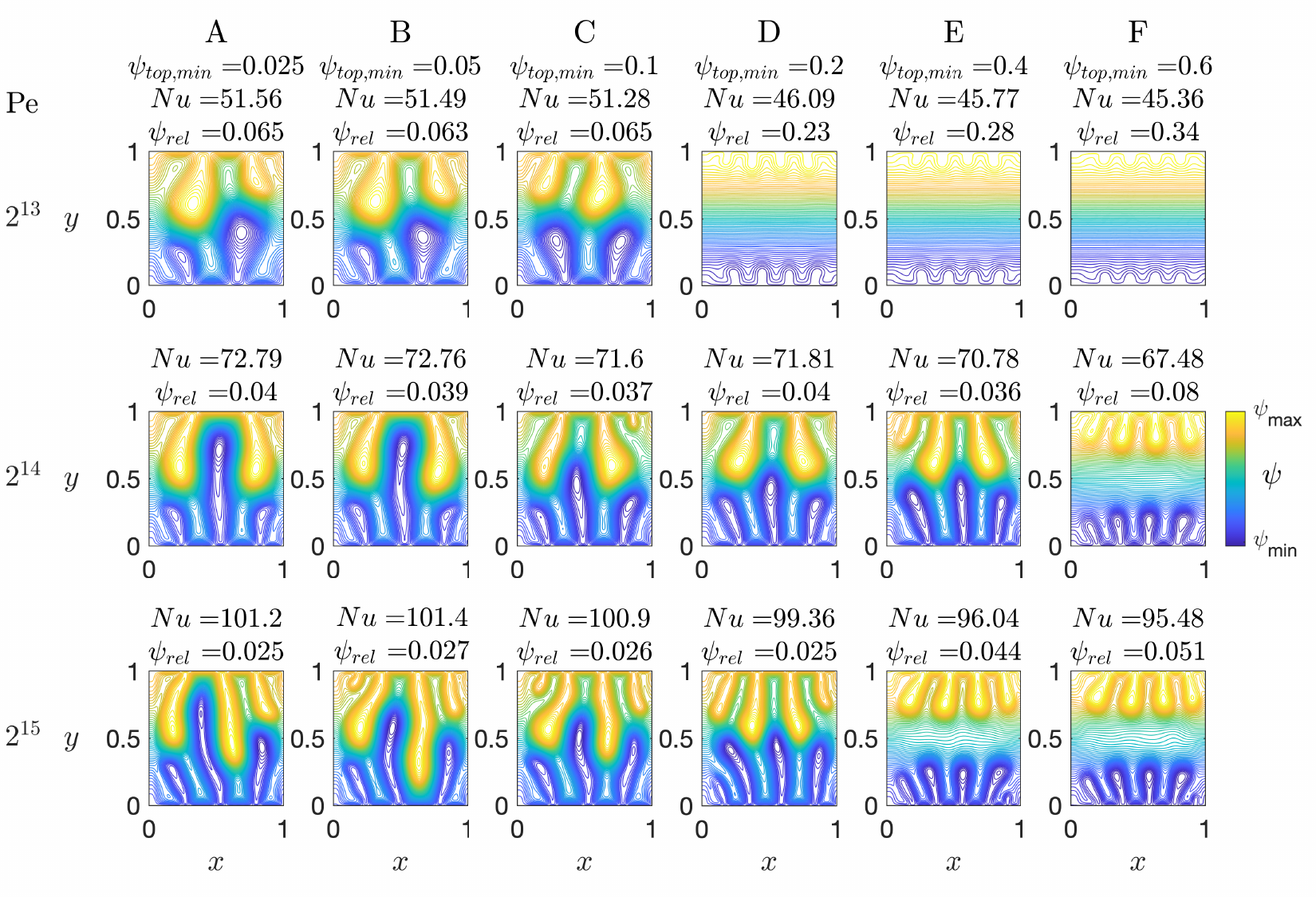}
    \caption{ We illustrate the effect of the $\psi_{top,min}$ parameter by showing contour plots of $\psi_{norm}$ for the optimal flows at $\psi_{top,min} = 0.025, 0.05, 0.1, 0.2, 0.4, 0.6$ with $Pe = 2^{13} - 2^{15}$. For all of the flows in this figure, we begin the optimization algorithm with 16 modes, end with 32 modes, and increase the number of modes by 1 every 20 iterations. These results were generated on a $256 \times 256$ mesh.}
    \label{fig:EffectofPTM}
\end{figure}

In figure~\ref{fig:EffectofPTM} we illustrate how varying the $\psi_{\text{top,min}}$ parameter influences the structure and performance of the optimal flows. This parameter only sets the {\it initial} strength of the Poiseuille component of the flow, as the $B$ coefficients are initially drawn from a Gaussian distribution with mean 0 and variance 1, resulting in initial magnitudes of order 1. During optimization, these coefficients are free to evolve---their magnitudes can grow or shrink without bound---and because the stream function is normalized in (\ref{Psi}), the actual strength of the Poiseuille flow can change freely, increasing or decreasing depending on the magnitudes of the $B$ coefficients. As $\psi_{top,min}$ is increased, the optimization is initially biased towards stronger Poiseuille-like flow, which appears to suppress the formation of finger-like structures that cause vertical mixing.  
This trend is illustrated in figure~\ref{fig:EffectofPTM}, where as $\psi_{\text{top,min}}$ increases (moving from left to right across each row), the fingers become progressively shorter and less pronounced, remaining closer to the wall. In each row, once $\psi_{\text{top,min}}$ exceeds a certain value, $\psi_{rel}$ (listed above each panel) jumps to a higher value. 
$\psi_{rel}$ represents the relative strength of the Poiseuille flow component, as explained in section \ref{SectionIVB1}.

Corresponding to the increase in $\psi_{rel}$ there is a noticeable drop in heat transfer performance. For instance, from column A to F there is a $12 \%$ decrease in $Nu$ for $Pe = 2^{13}$ in row 1, a $7 \% $ decrease for $Pe = 2^{14}$ in row 2, and a $6 \%$ decrease for $Pe = 2^{15}$ in row 3, indicating that the relative benefit of reducing $\psi_{\text{top,min}}$ is more pronounced at lower Péclet numbers. Structurally, we observe a distinct trend in the flow morphologies as $\psi_{\text{top,min}}$ is varied. Optimal flows at lower values of $\psi_{\text{top,min}}$ tend to develop fewer but longer finger-like protrusions that extend well into the center of the channel. In contrast, flows optimized with a higher $\psi_{\text{top,min}}$ display more numerous but shorter protrusions that remain closer to the walls. These shorter structures are less effective at promoting heat transport, resulting in lower overall Nusselt values.

\begin{figure}
\centering \includegraphics[width= 0.9 \linewidth]{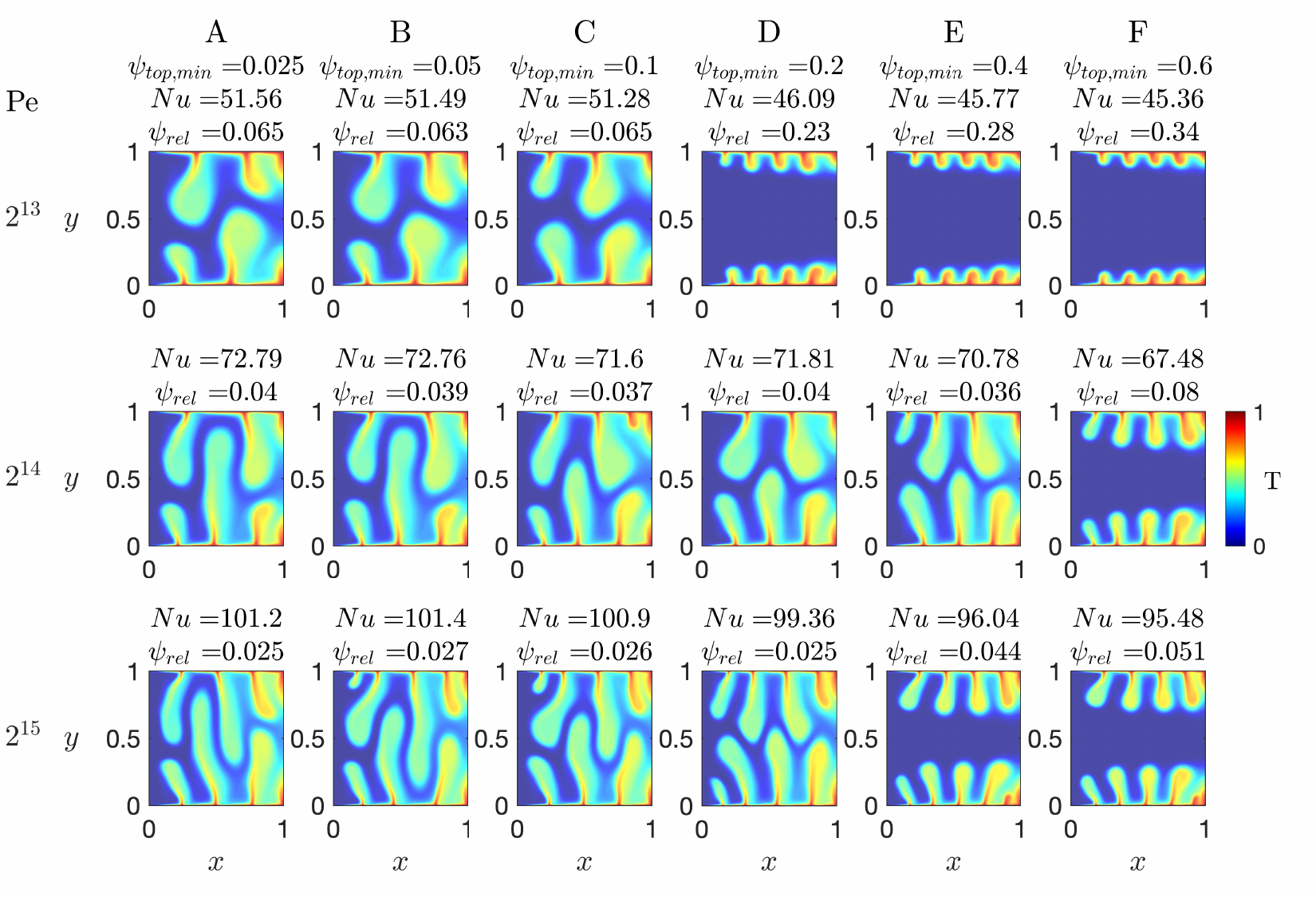}
    \caption{Temperature fields corresponding to the flows in figure \ref{fig:EffectofPTM}.}
    \label{fig:EffectofPTM_temp}
\end{figure}

Figure~\ref{fig:EffectofPTM_temp} shows the temperature fields corresponding to the flows in figure \ref{fig:EffectofPTM}. As in figure~\ref{fig:OptimalWavyFlowTemps}, we observe the presence of thermal fingers that closely follow the pattern of finger-like protrusions in figure~\ref{fig:EffectofPTM}. An interesting feature in figure~\ref{fig:EffectofPTM_temp} is that for larger values of $\psi_{\text{top,min}}$, such as in row 1 panels D $-$ F, the thermal fingers maintain higher temperatures over a greater portion of each finger.  The higher $\psi_{\text{top,min}}$ values lead to faster convective flow through the center of the channel, with warm fluid more confined near the top and bottom channel walls. Because the fingers are shorter, the fluid they contain has less surface area with which to exchange heat with the colder fluid in the middle of the channel. Nonetheless, the $Nu$ values are similar in magnitude from panels A to F.

\subsubsection{Estimated finger widths}
\label{SectionIVB6}

A key feature of the wavy flows is the formation of finger-like structures. Table~\ref{table:1} reports the average number of fingers emanating from the top and bottom walls for the 36 flows shown in figure~\ref{fig:OptimalWavyFlows} across a range of $Pe$ values. To quantify the characteristic lateral scale of these structures, we define an \textit{estimated finger width}, calculated as the domain length, 1, divided by the mean number of fingers. We assume that the fingers are approximately evenly distributed across the horizontal extent of the domain:
\begin{equation*}
    \text{Estimated Finger Width} = \frac{1}{\bar{n}_{f}}, \quad \text{where} \quad \bar{n}_{f} = \frac{n_{t} + n_{b}}{2}
\end{equation*}
where $n_t$ and $n_b$ denote the number of fingers extending from the top and bottom walls, respectively. This formula uses the approximate assumption that the fingers collectively span the full length of the channel.

\begin{table}
    \centering
\begin{tabular}{||c c c c c c c c||} 
 \hline\hline
  &$\bar{n}_{f}$& Opt Flow 1 & Opt Flow 2 & Opt Flow 3 & Opt Flow 4 & Opt Flow 5 & Opt Flow 6 \\ 
 \hline\hline
$Pe = 2^{12}$ \vline &  1.58 & 1 & 1.5 & 1.5& 1.5& 2& 2 \\ 
$Pe = 2^{13}$ \vline &  2.08 & 1.5 & 2 & 2& 2& 2.5 & 2.5 \\ 
$Pe = 2^{14}$ \vline&  2.58 & 2.5 & 2.5 & 2.5& 2& 3& 3 \\ 
$Pe = 2^{15}$ \vline&  3.33 & 3 & 3 & 3.5 & 3.5 & 3.5 & 3.5 \\ 
$Pe = 2^{16}$ \vline&  5.17 & 5 & 5 & 5 & 6 & 5 & 5 \\ 
$Pe = 2^{17}$ \vline&  6.42 & 6 & 6.5 & 7 & 6.5 & 6.5 & 6 \\ 
 \hline \hline
 \end{tabular}
 \caption{The average number of finger-like protrusions coming out of the top and bottom walls for each of the optimal flows in figure~\ref{fig:OptimalWavyFlows}. $\bar{n}_{f}$ is the average of average number of fingers for each $Pe = 2^{12} - 2^{17}$.}
 \label{table:1}
 \end{table}
 As $Pe$ increases from $2^{12}$ to $2^{17}$, $\bar{n}_{f}$ increases from $1.58$ to $6.42$, indicating a progressive narrowing and higher density of the fingers. The most significant jump occurs between $Pe = 2^{15}$ and $2^{16}$, where $\bar{n}_{f}$ jumps from 3.33 to 5.17. This increase coincides with the emergence of more bulbous fingers that exhibit stronger, larger, circular vortices at their tips, concentrated near the center of the channel. The jump in $\bar{n}_{f}$ may be partially due to this change in the flow morphology. At lower $Pe$, greater variability in finger number is observed across different optimal flows, reflecting a broader range of optimal flow structures. In contrast, at higher $Pe$, flows show more consistent finger counts, which suggest a convergence toward a more homogeneous set of flow configurations near the walls. Throughout the range from $Pe = 2^{12}$ to $2^{17}$, the estimated finger width decreases by approximately $75\%$, from about 0.63 to 0.16. As the fingers become thinner and more numerous with increasing $Pe$, they provide more pathways for the flow to carry warm fluid from the boundaries inward, increasing the heat flux at the boundaries. 

The narrowing of these fingers also sharpens horizontal temperature gradients between the hot finger cores and cooler surrounding fluid. This promotes more efficient horizontal diffusion of heat from the fingers into the interior of the channel.  Thus, even though the dominant temperature gradients and diffusive fluxes are horizontal near the finger edges, the overall heat transport is vertically oriented with hot fluid being convected into the channel from both boundaries. At higher $Pe$ values, the emergence of bulbous finger tips appears to utilize both vertical transport via the fingers as well as horizontal mixing through strong vortical structures aligned along channel centerline. These vortices help redistribute vertically convected heat more uniformly through the interior of the domain. The emergence of thinner and more numerous fingers increases the area of the interface between and hot and cold fluid as $Pe$ increases. 
\subsubsection{Scaling laws}
\label{SectionIVB7}
\begin{figure}
    \centering
    \includegraphics[width=1 \linewidth]{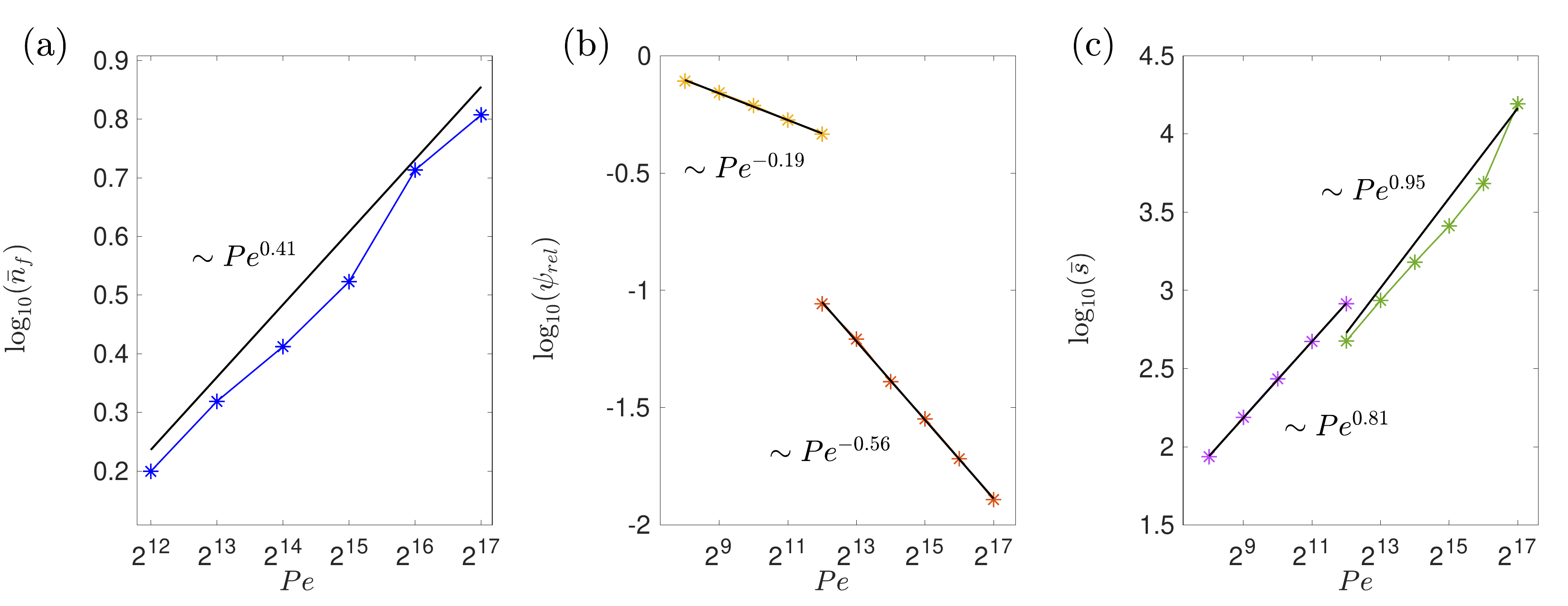}
    \caption{The average number of fingers, $(\bar{n}_{f})$, the relative background flow strength $(\psi_{rel})$, and the average maximum speed of the flow $(\bar{s})$ are plotted against $Pe$ on log scales. Each panel includes a linear regression fit in $
    \log-\log$ space, with the resulting power-law scalings labeled.} 
    \label{fig:CombinedComparisonPlots}
\end{figure}

Figure~\ref{fig:CombinedComparisonPlots} illustrates how the average number of fingers $\bar{n}_{f}$, $\psi_{rel}$, and the average maximum flow speeds scale with $Pe$. Panel (a) shows that $\bar{n}_{f} \sim Pe^{0.41}$, a sublinear increase in the number of fingers with $Pe$. As we’ve discussed, increasing $Pe$ leads to optimal flows characterized by thinner and more numerous fingers. At lower $Pe$, advection is moderately strong relative to diffusion, and fewer and wider fingers are more effective for distributing heat. At higher $Pe$, advection is stronger, and splitting into many narrow fingers is advantageous. Panel (b) shows that $\psi_{rel} \sim Pe^{-0.56}$ in the wavy flow regime. As $Pe$ increases, the relative flux through the channel decreases. More of the flow domain consists of slow or stagnant flow, while the exiting fluid reaches a higher temperature than at lower $Pe$. In panel (c) we show how the average maximum flow speed varies with $Pe$. We computed the flow speed throughout the domain, $s(x, y) = \sqrt{(u(x, y))^{2} + (v(x, y))^{2}}$, identified the maximum value of $s$ for each flow, and then averaged these maximum speeds across the six optima shown for
each $Pe$ value in figure \ref{fig:OptimalWavyFlows}. We observe a nearly linear scaling: $\bar{s} \sim Pe^{0.95}$, with a prefactor of about 0.1. 

\begin{figure}
    \centering
    \includegraphics[width=1\linewidth]{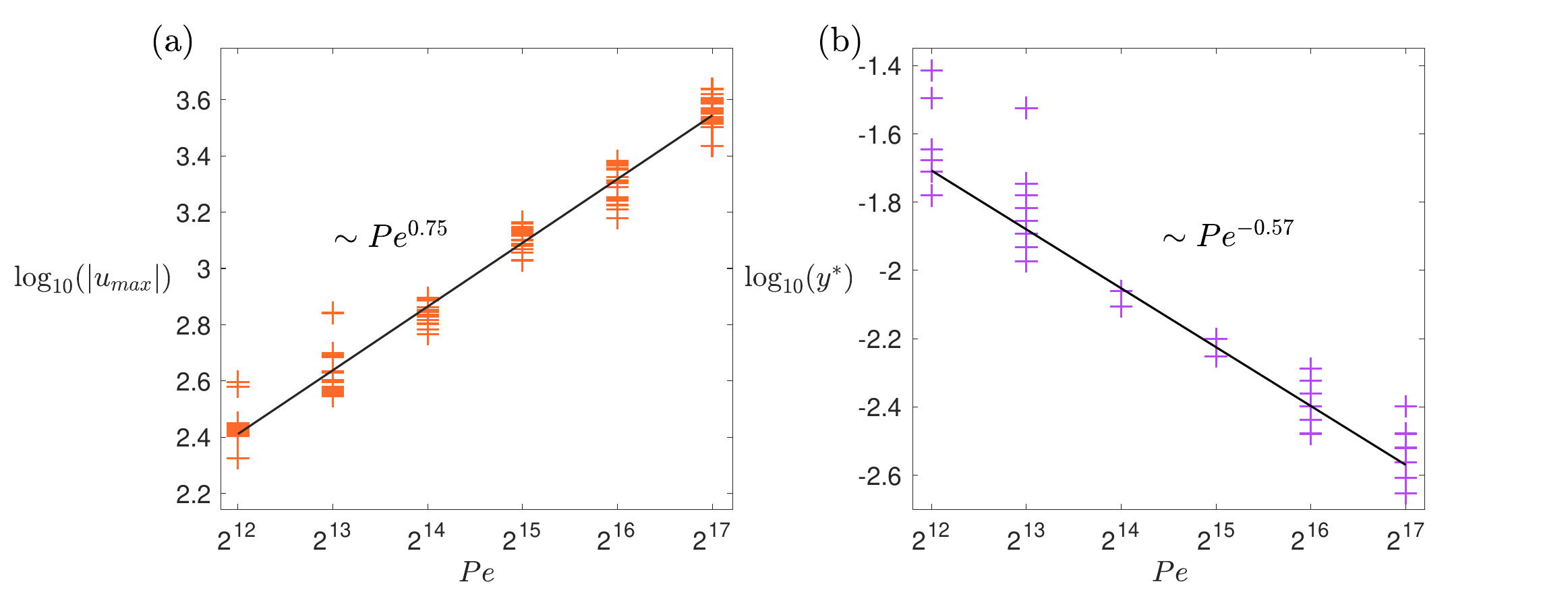}
    \caption{For the six optimal flows at each Pe, 24 data points for (a) $u_{max}$ and (b) $y^{*}$ in the boundary layers at six $Pe$.
    }
    \label{fig:CombinedScalingLaw}
\end{figure}
We also examined how the thicknesses of the velocity boundary layers at the walls scale with $Pe$. 
We estimate the boundary layer thickness at points on the walls with local maxima shear stress, i.e. at
$x^{*} = \text{arg max}_{\text{wall}} \Big(\Big|\frac{\partial u}{\partial y} \Big| \Big)$,
in order to avoid points where the boundary layer separates, which have zero shear stress, and where the boundary layer thickness is undefined.
At $x^{*}$, we identify the boundary layer thickness as the location of the maximum of horizontal velocity over a region near the wall, $y^{*} = \text{arg max}_{|y-y_{wall}| < 0.05}(u(x^{*}, y))$. We call the horizontal velocity there $u_{max} = u(x^{*}, y^*)$. We compute these quantities in the left and right halves of the channel, and at the top and bottom walls, for the six optimal flows at each $Pe$, and plot the results (24 points at each $Pe$) in figure \ref{fig:CombinedScalingLaw}. For the wavy flows, we find $u_{max} \sim Pe^{0.752}$, and $y^{*} \sim Pe^{-0.573}$. By contrast, for the unidirectional optima of figure \ref{fig:OptUnifFlows} and \cite{Alben2017Channel}, we had $u \sim Pe^{0.8}$, and $y^{*} \sim Pe^{-0.4}$. Thus, the wavy flows' boundary layers decrease more rapidly with $Pe$ than the unidirectional flows, while the velocities increase more slowly. This scaling of velocities is close to that found for the branched convection patterns in the 2D wall-to-wall optimal flows \cite{alben2023transition}.

In appendix \ref{AppendixB} we show the rescaled horizontal velocity profiles of the optimal wavy flows for $Pe$ ranging from $2^{12}$ to $2^{17}$ at the four sampling points. We see that a self-similar pattern emerges, but that there is considerably more variability than in the optimal unidirectional flow rescaling of figure \ref{fig:OptUnifFlows}(e). 
\begin{figure}
\centering
\includegraphics[width=1 \linewidth]{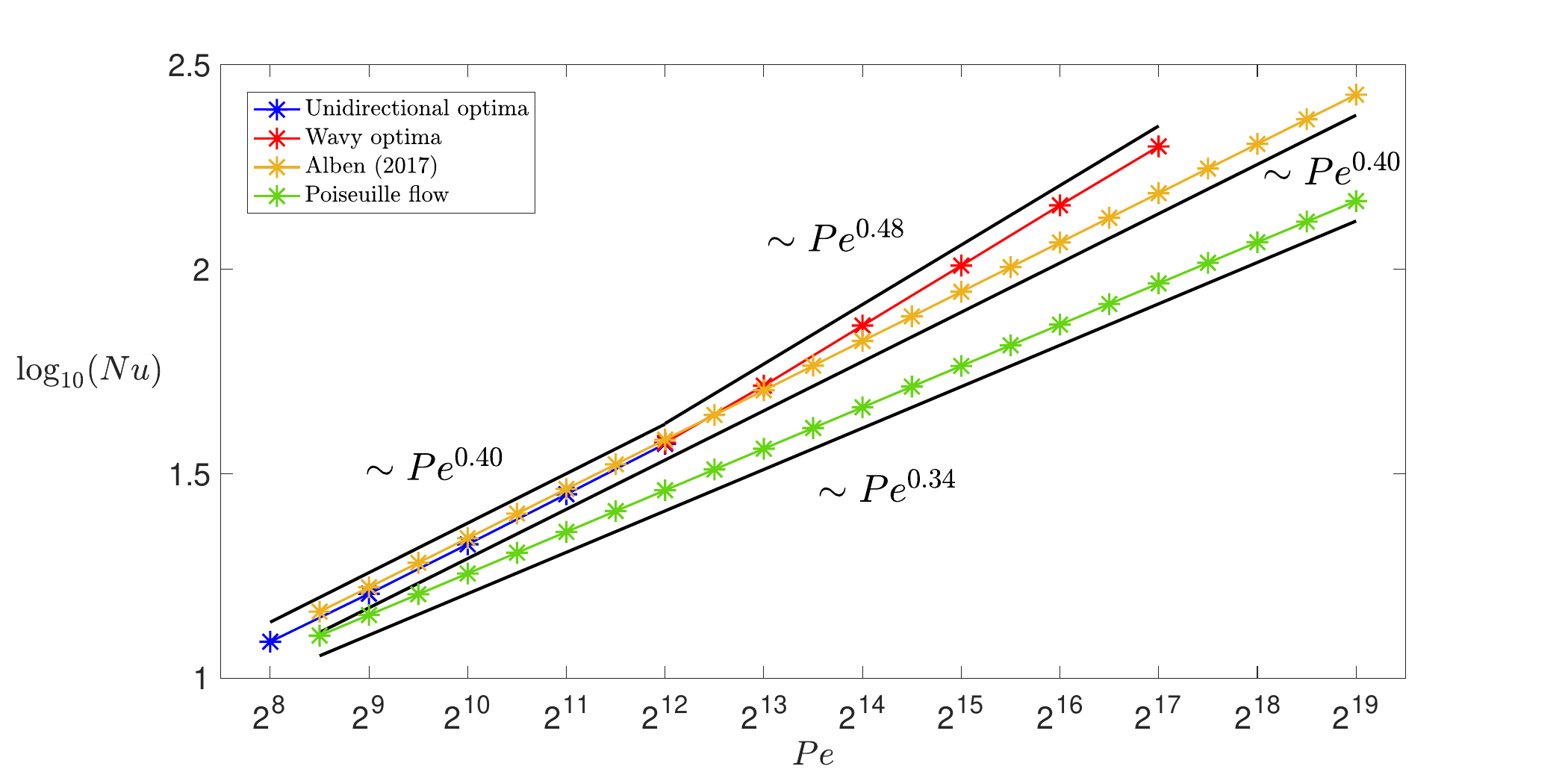}
    \caption{The top $Nu$ values at each $Pe = 2^{8} - 2^{17}$. The blue asterisked data are the top $Nu$ values at $Pe$ = $2^{8} - 2^{12}$ for the unidirectional optima in figure \ref{fig:OptUnifFlows}. The red asterisks are the top $Nu$ values for $Pe = 2^{12} - 2^{17}$ for the wavy optima in figure \ref{fig:OptimalWavyFlows}. The green asterisks are the $Nu$ values for Poiseuille flow, and the orange asterisks are the $Nu$ values for the unidirectional optima found in \cite{Alben2017Channel}.
    }
    \label{fig:NuvsPe}
\end{figure}
In figure~\ref{fig:NuvsPe} we show the $Nu$ values achieved by the unidirectional and wavy flows computed in this work, along with reference values for related flows. The blue asterisks represent the highest Nusselt values found for $Pe = 2^{8}$ to $2^{12}$, where approximately unidirectional flows were optimal. At $Pe = 2^{12}$, we show two data points, one for the best unidirectional flow which yields $Nu = 37.395$, and one for the best wavy flow which yields $Nu = 37.64$. Although the improvement is only modest at this transition point, $Pe = 2^{12}$ marks the onset of new wavy flow optima that outperform the unidirectional optima. The red asterisks represent the best $Nu$ values for $Pe = 2^{12}$ to $Pe = 2^{17}$, from wavy flows. These are compared to the orange asterisks, the previously best-known Nusselt values reported by \cite{Alben2017Channel}. The comparison reveals a clear and growing improvement in heat transfer performance by the wavy flow optima found in our study. The improvement reaches $30\%$ at $Pe = 2^{17}$. These progressively larger gains at higher Péclet numbers underscore the growing effectiveness of wavy flow structures in providing greater surface area for hot and cold fluid to exchange heat.

\section{Conclusions}
\label{SectionV}
We computed optimal flows for heat transfer in a channel using an adjoint-based optimization method and a mesh concentrated at the channel walls. We searched for optimal solutions within a class of flows composed of two components: a parabolic Poiseuille background flow and an interior component represented as a sum of products of Chebyshev polynomials. We varied the Péclet number from $2^{8} - 2^{17}$. For $Pe = 2^{8} - 2^{11}$, we found approximately unidirectional flows that aligned with previously found optima \cite{Alben2017Channel}. Beginning at $Pe = 2^{12}$, however, we discovered the emergence of wavy flows that outperformed unidirectional flows. These wavy optima exhibited finger-like protrusions that emanated from the top and bottom walls of the channels. As $Pe$ increased, the optimal flows had thinner and more numerous fingers, with the number of fingers proportional to $Pe^{0.41}$. Additionally, the boundary layer thickness scales approximately as $Pe^{-0.573}$, over which the flow reaches a maximum horizontal velocity $\sim Pe^{0.752}$. 

For $Pe = 2^{12} - 2^{17}$, our optimized flows achieved Nusselt values that scaled as $Pe^{0.48}$, improving upon the $Pe^{0.40}$ scaling reported by \cite{Alben2017Channel} for unidirectional flows. We posit that these wavy flows provide improved heat transfer due to their enhanced vertical and horizontal mixing and effective increase in surface area available for heat transfer from hot to cold fluid in the interior of the channel. Our work presents optimal flow patterns that may inspire improved engineering designs. Several questions remain open for future research. One is understanding the physical mechanisms that select the finger spacing and depth in the wavy optima. Additionally, it would be valuable to extend this analysis to even higher Péclet numbers, to explore whether additional flow structures emerge in that regime.

\section*{Acknowledgments}
This research was supported by the NSF-DMS Applied Mathematics program under
award number DMS-2204900.

\appendix
\section{Convergence studies}
We conducted convergence studies for the non-wavy and wavy optima shown in figures \ref{fig:OptUnifFlows} and \ref{fig:OptimalWavyFlows} for both horizontal and vertical mesh refinements at varying grid resolutions. For $Pe = 2^{8} - 2^{12}$, we computed results on a $128 \times 128$ mesh, and compared with $Nu$ values computed on meshes having $m = 64, 128, 256,$ and $512$ with $n = 128$ and meshes having $n = 64, 128, 256,$ and $512$ with $m = 128$. We found that the relative changes observed across all meshes and $Pe$ values were $\leq 1.5\% $. We then tested Nusselt values for optima with $Pe = 2^{8} - 2^{11}$, found on a $256 \times 256$ mesh, with meshes having $m = 128, 256, 512$ and $1024$ with $n = 256$ as well as $n = 128, 256, 512$, and $1024$ with $m = 256$.
\begin{table}
    \centering
     \begin{tabular}{|r||c|c|c|c|c|c|c|c|c||}
     \hline
     Flow & $m \times n $ &$Pe = 2^{12}$ & $\triangle$ & $Pe = 2^{13}$ & $\triangle$ & $Pe = 2^{14}$ & Error & $Pe = 2^{15}$ & $\triangle$  \\
     \hline\hline
     1 & $128 \times 256$ &$37.5644$ & -- & $51.7096$ & -- & $72.3653$ & -- & $101.1129$ & -- \\
     
     1 & $256 \times 256$ &$37.6439$ & $0.21\%$ & $51.8944$ & $0.36\%$& $72.7863$ & $0.58\%$& $101.9618$ & $0.84\%$ \\
    
     1 & $512 \times 256$ & $37.6607$ & $0.04\%$ & $51.9395$ & $0.09\%$ & $72.8961$ & $0.15 \%$& $102.1754$ & $0.21\%$ \\
    
     1 & $1024 \times 256$ & $37.6815$ & $0.06\%$ & $51.9819$ & $0.08\%$ & $72.9631$ & $0.09 \%$ & $102.2716$ & $0.09\%$ \\
    \hline \hline
     2 & $128 \times 256$ & $36.7321$ & -- & $51.4119$ & -- & $72.3648$ & -- & $100.8048$ & -- \\
     
     2 & $256 \times 256$ & $36.8143$ & $0.22\%$& $51.5617$ & $0.29 \%$& $72.7862$ & $0.58\%$& $101.6262$ & $0.81 \%$ \\
    
     2 & $512 \times 256$ &$36.8332$ & $0.05 \%$ & $51.6006$ & $0.08 \%$ & $72.8957$ & $0.15\%$& $101.8552$ & $0.23 \%$ \\
    
     2 & $1024 \times 256$ &$36.9236$ & $ 0.25 \%$ & $51.6487$ & $0.09\%$ & $72.9556$ & $0.08 \%$ & $101.9614$ & $0.10 \%$  \\
    \hline \hline
     3 & $128 \times 256$ & $36.6930$ & -- & $51.4161$ & -- & $71.9070$ & -- & $100.9640$ & -- \\
     
     3 & $256 \times 256$ & $36.7649$ & $0.20\%$ & $51.5587$ & $0.28 \%$& $72.2382$ & $0.46 \%$& $101.5429$ & $0.57 \%$ \\
    
     3 & $512 \times 256$ & $36.7811$ & $0.04\%$ & $51.5944$ & $0.07 \%$ & $72.3347$ & $0.13\%$& $101.7501$ & $0.20 \%$ \\
    
     3 & $1024 \times 256$ &$36.8414$ & $0.16 \%$ & $51.6510$ & $0.11 \%$ & $72.3956$ & $0.08 \%$ & $101.8545$ & $ 0.10 \%$  \\
    \hline \hline
     4 & $128 \times 256$ &$36.6862$ & -- & $51.1493$ & -- & $71.6186$ & -- & $100.8281$ & -- \\
     
     4 & $256 \times 256$ &$36.7600$ & $0.20 \%$& $51.3408$ & $0.37 \%$& $71.9543$ & $0.47 \%$& $101.5412$ & $0.71 \%$ \\
    
     4 & $512 \times 256$ &$36.7766$ & $0.05 \%$ & $51.3890$ & $ 0.09 \%$ & $72.0381$ & $0.12 \%$& $101.7605$ & $0.22 \%$ \\
    
     4 & $1024 \times 256$ &$36.8220$ & $0.12 \%$ & $51.4164$ & $0.05\%$ & $72.0970$ & $0.08 \%$ & $101.8554$ & $0.09\%$  \\
    \hline \hline
     5 & $128 \times 256$ &$36.2327$ & -- & $50.7121$ & -- & $71.6051$ & -- & $100.9356$ & -- \\
     
     5 & $256 \times 256$ &$36.3042$ & $0.20 \%$& $50.8799$ & $0.33\%$& $71.9476$ & $0.48 \%$& $101.4765$ & $0.54 \%$ \\
    
     5 & $512 \times 256$ &$36.3195$ & $ 0.04 \%$ & $50.9192$ & $0.08 \%$ & $72.0514$ & $0.14\%$& $101.6935$ & $0.21\%$ \\
    
     5 & $1024 \times 256$ &$36.3887$ & $ 0.19 \%$ & $50.9604$ & $0.08 \%$ & $72.1102$ & $0.08\%$ & $101.7809$ & $0.09\%$  \\
    \hline \hline
     6 & $128 \times 256$ & $36.1986$ & -- & $50.6048$ & -- & $71.1763$ & -- & $100.8205$ & -- \\
     
     6 & $256 \times 256$ &$36.2811$ & $ 0.23 \%$& $50.7931$ & $0.37 \%$& $71.5052$ & $0.46 \%$& $101.3547$ & $0.53 \%$ \\
    
     6 & $512 \times 256$ &$36.3000$ & $ 0.05 \%$ & $50.8408$ & $0.09\%$ & $71.5874$ & $0.11 \%$& $101.5276$ & $0.17 \%$ \\
    
     6 & $1024 \times 256$ &$36.3316$ & $0.09 \%$ & $50.9012$ & $0.12\%$ & $71.6484$ & $0.09\%$ & $101.6205$ & $0.09 \%$  \\
    \hline \hline
     \end{tabular} \label{AppendixA}
 \caption{We present a convergence study for the optimal flows shown in figure \ref{fig:OptimalWavyFlows} for $Pe = 2^{12} - 2^{15}$. These optima were computed on a $256 \times 256$ mesh, and we compare with results computed on meshes of size $m = 128, 256, 512$, and $1024 \times n = 256$. We give the relative percentage change, $\triangle$, when comparing $Nu$ computed on a coarser mesh with $Nu$ computed using the next finer mesh.}
 \label{table:2}
    \end{table}
In these tests, the relative changes were $\leq 5.7 \%$. The data show that the non-wavy flows, optimal for $Pe = 2^{8} - 2^{11}$, had better convergence and higher $Nu$ values on the coarser $128 \times 128$ mesh, while the wavy flows (which begin to outperform the non-wavy ones at $Pe = 2^{12}$) exhibited noticeably better convergence on the finer $256 \times 256$ mesh. One possible reason for this difference is that the non-wavy flows, being simpler in structure, can be represented with fewer modes and can thus be resolved with a coarser mesh, resulting in better-conditioned matrices, which can improve the accuracy of computing $Nu$ and its gradient. Also, with fewer modes the optimization search is conducted in a space of lower dimension, which may speed up convergence to a local minimizer.
In contrast, the wavy flows are more complex, so more modes and finer meshes are needed to represent and resolve them.

Tables \ref{table:2} and \ref{table:3} show a comparison of $Nu$ values for the optima presented in figure \ref{fig:OptimalWavyFlows} (computed on a $256 \times 256$ mesh) with meshes where $m = 128, 256, 512$ and $1024$, and $n = 256$ as well as with meshes having $n = 128, 256, 512$, and $1024$ with $m = 256$ respectively.
 \begin{table}
    \centering
     \begin{tabular}{|r||c|c|c|c|c|c|c|c|c||}
     \hline
     Flow & $m \times n $ &$Pe = 2^{12}$ & $\triangle$ & $Pe = 2^{13}$ & $\triangle$ & $Pe = 2^{14}$ & Error & $Pe = 2^{15}$ & $\triangle$  \\
     \hline\hline
     1 & $256 \times 128$ &$37.6922$ & -- & $51.9942$ & -- & $73.0478$ & -- & $102.4948$ & -- \\
     
     1 & $256 \times 256$ & $37.6439$ & $0.13 \%$ & $51.8944$ & $0.19 \%$& $72.7863$ & $0.36 \%$& $101.9618$ & $0.52 \%$ \\
    
     1 & $256 \times 512$ & $37.6446$ & $0.002 \%$ & $51.8949$ & $0.0008 \%$ & $72.7661$ & $0.03 \%$& $101.9195$ & $0.04 \%$ \\
    
     1 & $256 \times 1024$ & $37.6809$ & $0.10 \%$ & $51.9123$ & $0.03 \%$ & $72.7623$ & $0.005 \%$ & $101.9253$ & $0.006 \%$ \\
    \hline \hline
     2 & $256 \times 128$ & $36.8734$ & -- & $51.6929$ & -- & $73.0466$ & -- & $102.1828$ & -- \\
     
     2 & $256 \times 256$ & $36.8143$ & $0.16 \%$ & $51.5617$ & $0.25 \%$& $72.7862$ & $0.36 \%$& $101.6262$ & $0.54 \%$ \\
    
     2 & $256 \times 512$ & $36.8188$ & $0.01 \%$ & $51.5595$ & $0.004 \%$ & $72.7645$ & $0.03 \%$& $101.5776$ & $0.05 \%$ \\
    
     2 & $256 \times 1024$ & $36.8149$ & $0.01 \%$ & $51.4884$ & $0.14 \%$ & $72.7759$ & $0.02 \%$ & $101.5567$ & $0.02 \%$ \\
    \hline \hline
     3 & $256 \times 128$ &$36.8225$ & -- & $51.6884$ & -- & $72.4965$ & -- & $102.3543$ & -- \\
     
     3 & $256 \times 256$ & $36.7649$ & $0.16 \%$ & $51.5587$ & $0.25 \%$& $72.2382$ & $0.36 \%$& $101.5429$ & $0.79 \%$ \\
    
     3 & $256 \times 512$ & $36.7579$ & $0.02 \%$ & $51.5469$ & $0.02 \%$ & $72.2214$ & $0.02 \%$& $101.4616$ & $0.08 \%$ \\
    
     3 & $256 \times 1024$ & $36.8067$ & $0.13 \%$ & $51.5500$ & $0.006 \%$ & $72.1907$ & $0.04 \%$ & $101.4322$ & $0.03 \%$ \\
    \hline \hline
     4 & $256 \times 128$ & $36.8222$ & -- & $51.4946$ & -- & $72.2602$ & -- & $102.1673$ & -- \\
     
     4 & $256 \times 256$ & $36.7600$ & $ 0.17 \%$ & $51.3408$ & $0.30 \%$& $71.9543$ & $0.42 \%$& $101.5412$ & $0.61 \%$ \\
    
     4 & $256 \times 512$ & $36.7584$ & $ 0.004 \%$ & $51.3239$ & $0.03 \%$ & $71.9271$ & $ 0.04 \%$& $101.4760$ & $0.06 \%$ \\
    
     4 & $256 \times 1024$ & $36.7897$ & $0.09 \%$ & $51.2844$ & $0.08 \%$ & $71.8991$ & $0.04 \%$ & $101.4598$ & $0.02 \%$ \\
    \hline \hline
     5 & $256 \times 128$ &$36.3756$ & -- & $51.0401$ & -- & $72.2406$ & -- & $102.2476$ & -- \\
     
     5 & $256 \times 256$ & $36.3042$ & $0.20 \%$ & $50.8799$ & $0.31 \%$& $71.9476$ & $0.41 \%$& $101.4765$ & $0.75 \%$ \\
    
     5 & $256 \times 512$ & $36.2928$ & $0.03 \%$ & $50.8570$ & $0.05 \%$ & $71.9219$ & $0.04 \%$& $101.4033$ & $0.07 \%$ \\
    
     5 & $256 \times 1024$ & $36.3912$ & $ 0.27 \%$ & $50.8639$ & $0.01 \%$ & $71.9396$ & $0.02 \%$ & $101.4243$ & $0.02 \%$ \\
    \hline \hline
     6 & $256 \times 128$ &$36.3550$ & -- & $50.9340$ & -- & $71.7911$ & -- & $101.9810$ & -- \\
     
     6 & $256 \times 256$ & $36.2811$ & $0.20 \%$ & $50.7931$ & $0.28 \%$& $71.5052$ & $0.40 \%$& $101.3547$ & $0.61 \%$ \\
    
     6 & $256 \times 512$ & $36.2828$ & $0.005 \%$ & $50.7787$ & $0.03 \%$ & $71.4797$ & $0.04 \%$& $101.2985$ & $0.06 \%$ \\
    
     6 & $256 \times 1024$ & $36.2776$ & $0.01 \%$ & $50.8074$ & $0.06 \%$ & $71.4615$ & $0.03 \%$ & $101.2576$ & $0.04 \%$ \\
    \hline \hline
     \end{tabular}
 \caption{We present a convergence study for the optimal flows shown in figure \ref{fig:OptimalWavyFlows} for $Pe = 2^{12} - 2^{15}$. These optima were computed on a $256 \times 256$ mesh, and we compare with results computed on meshes of size $m = 256 \times n = 128, 256, 512$, and $1024$. We give the relative percentage change, $\triangle$, when comparing $Nu$ computed on a coarser mesh with $Nu$ computed using the next finer mesh.}
 \label{table:3}
 \end{table}
When $m$ is varied, these flows yield consistent Nusselt numbers where the relative percent change, $ \triangle = \frac{Nu_{\text{finer}} - Nu_{\text{coarser}}}{Nu_{\text{coarser}}} \times 100\%$, is $ \leq 0.84\% $ across all meshes. Furthermore, we observe that the relative changes decrease monotonically in most cases, indicating that the $Nu$ values are converging as the mesh is refined. While small variations in the percent changes may arise from flow-specific characteristics, the convergence trend generally holds. Across all flow cases and Péclet numbers, the relative percent change from the $256 \times 256$ mesh to the $512 \times 256$ mesh ranges from 0.04\% to 0.23\%, with most values falling between 0.08\% and 0.15\%. Additionally, from $256 \times 256$ to $1024 \times 256$, the total relative change is generally under $0.3 - 0.4 \%$. These small differences suggest that the $256 \times 256$ mesh is sufficiently fine for accurate numerical solutions. We see that the Nusselt number increases slightly with mesh refinement, but the change tapers off quickly after $256 \times 256$, indicating that the solution is approaching asymptotic convergence, and the $256 \times 256$ mesh is already near the plateau. 
\begin{table}
    \centering
     \begin{tabular}{|r||c|c|c|c|c||}
     \hline
     Flow & $m \times n $ &$Pe = 2^{16}$ & $\triangle$ & $Pe = 2^{17}$ & $\triangle$  \\
     \hline\hline
     1 & $192 \times 384$ & $142.4656$ & -- & $205.3866$ & --  \\
     
     1 & $384 \times 384$ & $143.1550$ & $0.48 \%$ & $199.4576$ & $2.89 \%$  \\
    
     1 & $768 \times 384$ & $143.3720$ & $0.15 \%$ & $197.5282$ & $0.97 \%$  \\
    
     1 & $1536 \times 384$ &$143.4412$ & $0.05 \%$ & $197.5706$ & $0.02 \%$  \\
    \hline \hline
     2 & $192 \times 384$ & $142.4828$ & -- & $202.7921$ & --  \\
     
     2 & $384 \times 384$ & $143.0766$ & $0.42 \%$ & $198.9391$ & $1.90 \%$  \\
    
     2 & $768 \times 384$ & $143.2412$ & $0.12 \%$ & $192.9612$ & $3.00 \%$  \\
    
     2 & $1536 \times 384$ &$143.3397$ & $0.07 \%$ & $193.2839$ & $0.17 \%$  \\
    \hline \hline
     3 & $192 \times 384$ & $142.5710$ & -- & $204.1937$ & --  \\
     
     3 & $384 \times 384$ & $142.7600$ & $0.13 \%$ & $198.4390$ & $2.82 \%$  \\
    
     3 & $768 \times 384$ & $142.8391$ & $0.06 \%$ & $189.4507$ & $4.53 \%$  \\
    
     3 & $1536 \times 384$ &$142.9425$ & $0.07 \%$ & $189.5077$ & $0.03 \%$  \\
    \hline \hline
     4 & $192 \times 384$ & $141.4018$ & -- & $204.9158$ & --  \\
     
     4 & $384 \times 384$ & $142.6187$ & $0.86 \%$ & $197.7686$ & $3.49 \%$  \\
    
     4 & $768 \times 384$ & $142.8289$ & $0.15 \%$ & $190.8925$ & $3.48 \%$  \\
    
     4 & $1536 \times 384$ &$142.9987$ & $0.12 \%$ & $190.7462$ & $0.08 \%$  \\
    \hline \hline
     5 & $192 \times 384$ & $142.0163$ & -- & $207.1616$ & --  \\
     
     5 & $384 \times 384$ & $142.5507$ & $0.38 \%$ & $197.6423$ & $4.60 \%$  \\
    
     5 & $768 \times 384$ & $142.7357$ & $0.13 \%$ & $193.3221$ & $2.19 \%$  \\
    
     5 & $1536 \times 384$ &$142.8095$ & $0.05 \%$ & $193.0329$ & $0.15 \%$  \\
    \hline \hline
     6 & $192 \times 384$ & $140.6087$ & -- & $205.1411$ & --  \\
     
     6 & $384 \times 384$ & $141.7073$ & $0.78 \%$ & $197.5330$ & $3.71 \%$  \\
    
     6 & $768 \times 384$ & $141.9287$ & $0.16 \%$ & $194.7808$ & $1.39 \%$  \\
    
     6 & $1536 \times 384$ &$142.0347$ & $0.07 \%$ & $194.5448$ & $0.12 \%$  \\
    \hline \hline
     \end{tabular}
 \caption{We present a convergence study for the optimal flows shown in figure \ref{fig:OptimalWavyFlows} for $Pe = 2^{16} - 2^{17}$. These optima were computed on a $384 \times 384$ mesh, and we compare with results computed on meshes of size $m = 192, 384, 768$, and $1536 \times n = 384$. We give the relative percentage change, $\triangle$, when comparing $Nu$ computed on a coarser mesh with $Nu$ computed using the next finer mesh.}
 \label{table:4}
 \end{table}
 
 \begin{table}
    \centering
     \begin{tabular}{|r||c|c|c|c|c||}
     \hline
     Flow & $m \times n $ &$Pe = 2^{16}$ & $\triangle$ & $Pe = 2^{17}$ & $\triangle$  \\
     \hline\hline
     1 & $384 \times 192$ &$143.9668$ & -- & $200.4805$ & --  \\
     
     1 & $384 \times 384$ & $143.1550$ & $0.56 \%$ & $199.4576$ & $0.51 \%$  \\
    
     1 & $384 \times 768$ & $142.9968$ & $0.11 \%$ & $199.2540$ & $0.10 \%$  \\
    
     1 & $384 \times 1536$ &$143.0183$ & $0.02 \%$ & $199.2628$ & $0.004 \%$  \\
    \hline \hline
     2 & $384 \times 192$ &$143.5301$ & -- & $196.4816$ & --  \\
     
     2 & $384 \times 384$ & $143.0766$ & $0.32 \%$ & $198.9391$ & $1.25 \%$  \\
    
     2 & $384 \times 768$ & $142.9859$ & $0.06 \%$ & $199.6031$ & $0.33 \%$  \\
    
     2 & $384 \times 1536$ &$142.9954$ & $0.007 \%$ & $199.8039$ & $0.10 \%$  \\
    \hline \hline
     3 & $384 \times 192$ &$143.4382$ & -- & $196.4575$ & --  \\
     
     3 & $384 \times 384$ & $142.7600$ & $0.47 \%$ & $198.4390$ & $1.01 \%$  \\
    
     3 & $384 \times 768$ & $142.6167$ & $0.10 \%$ & $198.8270$ & $0.20 \%$  \\
    
     3 & $384 \times 1536$ & $142.6291$ & $0.009 \%$ & $199.0162$ & $0.10 \%$  \\
    \hline \hline
     4 & $384 \times 192$ & $143.1875$ & -- & $195.5725$ & --  \\
     
     4 & $384 \times 384$ & $142.6187$ & $0.40 \%$ & $197.7686$ & $1.12 \%$  \\
    
     4 & $384 \times 768$ & $142.4922$ & $0.09 \%$ & $198.4847$ & $0.36 \%$  \\
    
     4 & $384 \times 1536$ &$142.4858$ & $0.005 \%$ & $198.6861$ & $0.10 \%$  \\
    \hline \hline
     5 & $384 \times 192$ & $143.3809$ & -- & $197.5020$ & --  \\
     
     5 & $384 \times 384$ & $142.5507$ & $0.58 \%$ & $197.6423$ & $0.07 \%$  \\
    
     5 & $384 \times 768$ & $142.4102$ & $0.10 \%$ & $197.7540$ & $0.06 \%$  \\
    
     5 & $384 \times 1536$ & $142.4085$ & $0.001 \%$ & $197.8733$ & $0.06 \%$  \\
    \hline \hline
     6 & $384 \times 192$ &$142.2114$ & -- & $198.2569$ & --  \\
     
     6 & $384 \times 384$ & $141.7073$ & $0.35 \%$ & $197.5330$ & $0.37 \%$  \\
    
     6 & $384 \times 768$ &$141.6039$ & $0.07 \%$ & $197.4024$ & $0.07 \%$  \\
    
     6 & $384 \times 1536$ &$141.5868$ & $0.01 \%$ & $197.4101$ & $0.004 \%$  \\
    \hline \hline
    
     \end{tabular}
 \caption{We present a convergence study for the optimal flows shown in figure \ref{fig:OptimalWavyFlows} for $Pe = 2^{16} - 2^{17}$. These optima were computed on a $384 \times 384$ mesh, and we compare with results computed on meshes of size $m = 384 \times n = 192, 384, 768$, and $1536$. We give the relative percentage change, $\triangle$, when comparing $Nu$ computed on a coarser mesh with $Nu$ computed using the next finer mesh.}
 \label{table: 5}
 \end{table}
 Table \ref{table:3} shows a similar pattern where the relative percent change is $\leq 0.79 \%$ across all meshes. When varying $n$, the relative percent change from the $256 \times 256$ mesh to the $512 \times 256$ mesh ranges from 0.008\% to 0.08\%, with most values typically under $0.05 \%$. Further refinement to $256 \times 1024$ yields differences below $0.1 \%$, often near $0.01 - 0.03\%$. The smaller changes at the finer meshes, beyond $256 \times 256$, indicate that the solution is well-resolved. The finer meshes offer minimal accuracy gains relative to their increased computational costs. In the $256 \times 256$ mesh study, we found that convergence worsened at $Pe = 2^{16}$ with relative change being $\leq 5.6 \%$. This led us to test a $384 \times 384$ mesh. In tables \ref{table:4} and \ref{table: 5} we compare $Nu$ values computed for flows with $Pe = 2^{16} - 2^{17}$ on a $384 \times 384$ mesh against meshes having $m = 192, 384,768$ and $1536$ with $n = 384$ and $n = 192, 384, 768$ and $1536$ with $m = 384$ respectively. We observed better convergence for $Pe = 2^{16}$ on the $384 \times 384$ mesh, which motivated the use of a finer mesh at these higher Péclet values. For instance, in table \ref{table:4} the relative change across all meshes is $ \leq 0.86 \%$ at $Pe = 2^{16}$, a significantly better upper bound and the percent changes decrease with further refinement in most cases. In table \ref{table:4}, between $384 \times 384$ to $768 \times 384$, the relative percent changes are around $0.1 - 0.2 \%$ for $Pe = 2^{16}$ and $1 - 5\%$ for $Pe = 2^{17}$. From $768 \times 384$ to $1536 \times 384$, they drop below $0.2 \%$. A similar trend is observed in table \ref{table: 5}. Moving from $384 \times 192$ to $384 \times 384$ yields changes up to $0.58 \%$ in $Nu$. As in the horizontal refinement study, further refinement to $384 \times 768$ and $384 \times 1536$ yields progressively smaller changes, often under $0.1 \%$, and in many cases below $0.01 \%$. Together, these results demonstrate that the $384 \times 384$ mesh captures nearly all of the relevant behavior of the optimal flows for $Pe = 2^{16} - 2^{17}$, with improvements from further refinement becoming marginal. Across all flows, and both $Pe$ values, the relative changes in $Nu$ beyond the $384 \times 384$ mesh are typically below $0.36 \%$. We observe some larger relative changes in the refinement from $384 \times 384$ to $768 \times 384$ for $Pe = 2^{17}$, between $1 - 5 \%$. However, most of these cases show the percent changes decreasing, with only 2 cases (optimal flows 2 and 3) showing an increase in the percent change. Although the exact cause is not fully understood, separate tests examining the stretching factor $\lambda$ revealed a trade-off between mesh resolution and matrix condition number. Coarser meshes generally yield better-conditioned systems but offer lower spatial resolution, while finer meshes provide greater resolution at the cost of a worse condition number. In cases where coarser meshes produce better results, we suspect that the improved conditioning plays a more significant role in identifying better optima. Conversely, when finer meshes exhibit poorer convergence, we believe that the increased condition number leads to larger relative percent changes, even though the optimal flow structures themselves remain well-resolved.

\section{Rescaled horizontal velocity profiles} 

\begin{figure}[b]
\hspace{-0.75in}
\includegraphics[width=1.1\linewidth]{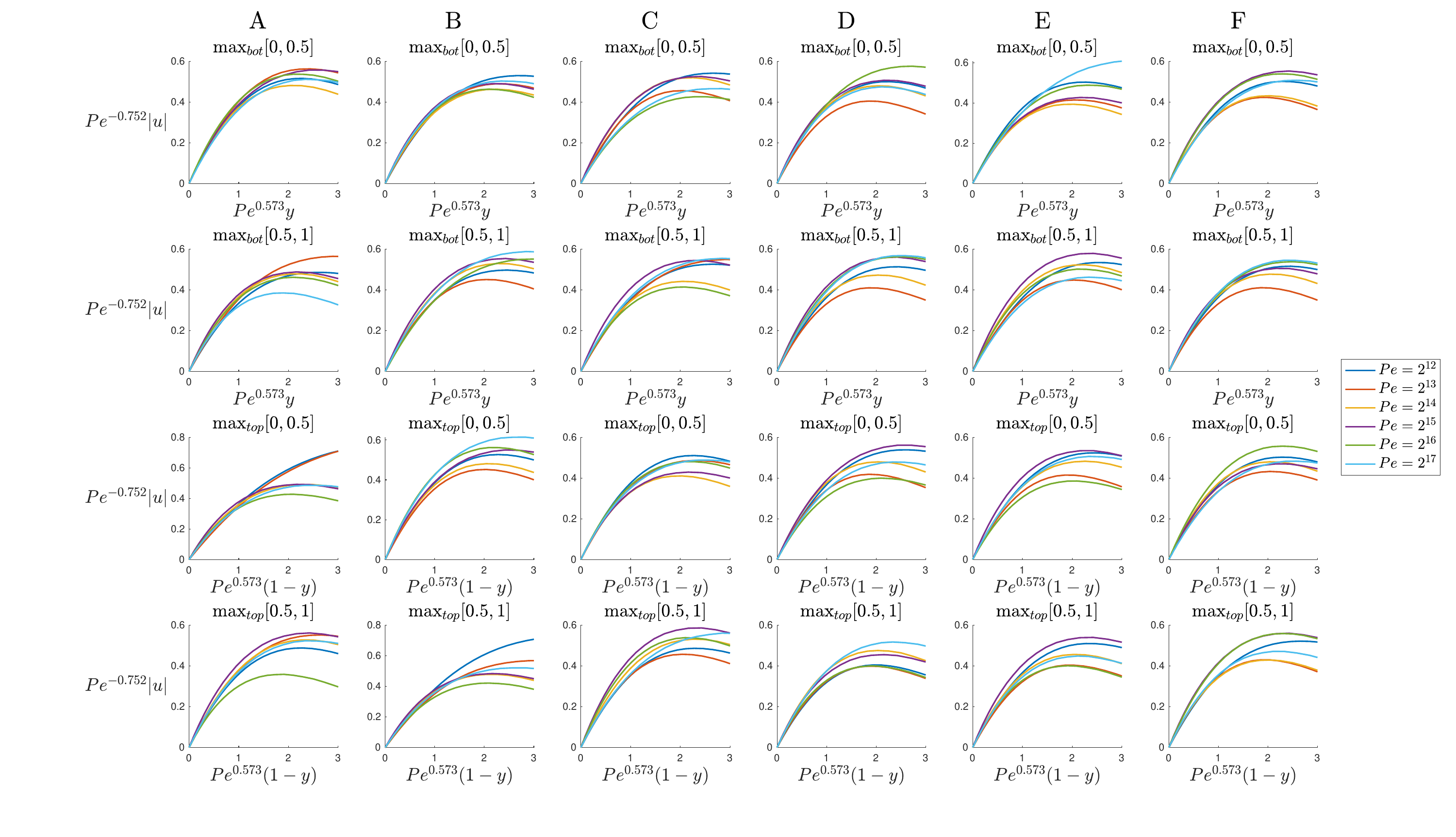}
    \caption{The rescaled horizontal velocity profile plots with for all 36 flows in figure \ref{fig:OptimalWavyFlows} with the scaling laws ($u_{max} \sim Pe^{0.752}$, and $y^{*} \sim Pe^{-0.573}$) found in figure \ref{fig:CombinedScalingLaw} for $Pe$ values ranging from $2^{12} - 2^{17}$.}\label{fig:selfsimilarplot}
\end{figure}\label{AppendixB}

In figure \ref{fig:selfsimilarplot}, we present the plots of the rescaled horizontal velocity profiles for all the Péclet values at the four $x^{*}$ values calculated in section \ref{SectionIVB7} 
for all 36 flows from figure \ref{fig:OptimalWavyFlows} rescaled with the scaling laws found in figure \ref{fig:CombinedScalingLaw}. We can see that the profiles overlap less than in figure \ref{fig:OptUnifFlows}(e) but still show good agreement. The reduced overlap may stem from the greater complexity of the fingering flows, which exhibit substantial variation in both horizontal and vertical directions. This increased spatial heterogeneity particularly in regions with strong curvature, flow reversal, or intricate finger structures can hinder the effectiveness of rescaling, making perfect collapse of the velocity profiles more difficult to achieve.

\section{Mode Initialization and Update Schedule for BFGS Optimization}
\label{AppendixC}
In table \ref{tableVI}, we present the starting modes, ending modes, and update frequency for each of the 36 flows shown in figure \ref{fig:OptimalWavyFlows}. The starting modes refer to the number of modes used to initialize the BFGS optimization. The ending modes represent the total number of modes at convergence. The update frequency indicates how often the mode count was incremented by one (i.e. every 2, 5, 10, or 20 iterations), until the predetermined number of ending modes is reached. 
\begin{table}[b]
    \centering
    \begin{tabular}{|r||c|c|c|c|c|c|c||} 
     \hline
     $Pe$ &  & Opt Flow 1 & Opt Flow 2 & Opt Flow 3  & Opt Flow 4  & Opt Flow 5 & Opt Flow 6 \\
     \hline\hline
     & Starting Modes & 24 & 16 & 28 & 28 & 28  & 28\\
     $2^{12} $ & Ending Modes & 32 & 32 & 32 & 32 & 32 & 32 \\
      & Update Frequency & 20 & 20 & 10 & 20 & 20 & 2\\
     \hline \hline
     & Starting Modes & 24 & 16 & 24 & 24 & 28 & 28\\
     $2^{13} $ & Ending Modes & 32 & 32 & 32 & 32 & 32 & 32 \\
      & Update Frequency & 5 & 20 & 5 & 20 & 10 & 2\\
     \hline \hline
     & Starting Modes & 28 & 16 & 16 & 24 & 28 & 28 \\
     $2^{14} $ & Ending Modes & 32 & 32 & 32 & 32 & 32 & 32 \\
      & Update Frequency & 20 & 20 & 10 & 5 & 20 & 20\\
     \hline \hline
     & Starting Modes & 28 & 28 & 16 & 16 & 24 & 24\\
     $2^{15} $ & Ending Modes & 32 & 32 & 32 & 32 & 32 & 32 \\
      & Update Frequency & 20 & 20 & 2 & 5 & 10 & 20 \\
     \hline \hline
     & Starting Modes & 42 & 24 & 42 & 42 & 42 & 24 \\
     $2^{16} $ & Ending Modes & 48 & 48 & 48 & 48 & 48 & 48 \\
      & Update Frequency & 5 & 10 & 2 & 2 & 10 & 10 \\
     \hline \hline
     & Starting Modes & 36 & 42 & 36 & 42 & 42 & 36\\
     $2^{17} $ & Ending Modes & 60 & 48 & 48 & 60 & 60 & 60 \\
      & Update Frequency & 10 & 20 & 5 & 2 & 5 & 20\\
     \hline \hline
     \end{tabular}
    \caption{The mode initialization and update strategy for all 36 optimal flows presented in figure \ref{fig:OptimalWavyFlows} for $Pe = 2^{12} - 2^{17}$.}
    \label{tableVI}
\end{table} 
\section{Additional Optimal Flows for $\lambda = 0.97$}
\label{AppendixD}

\begin{figure}[h]
    \centering \includegraphics[width=0.85\linewidth]{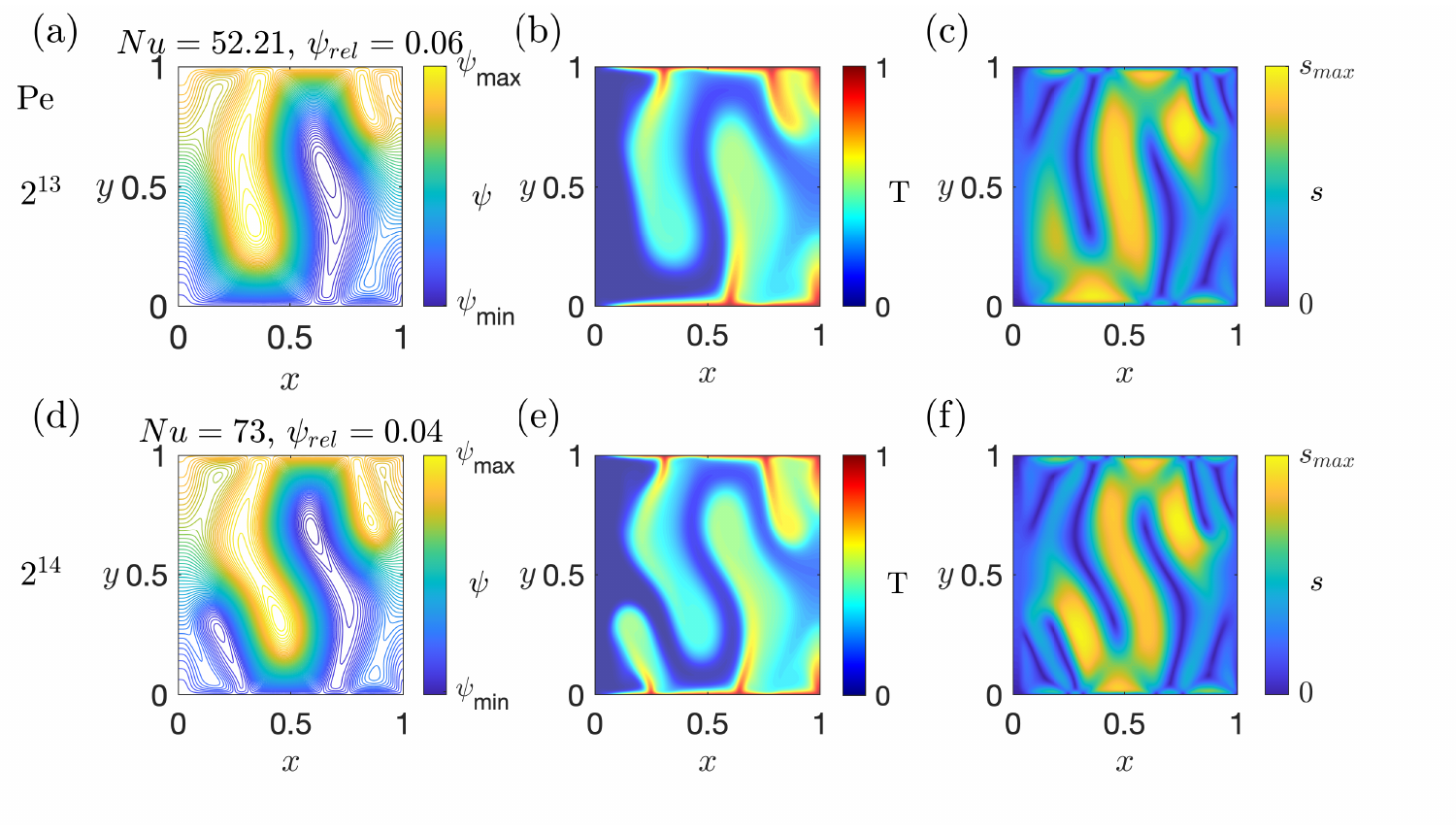}
    \caption{
    Stream function contour plots ((a) and (d)), temperature fields ((b) and (e)), and speed plots ((c) and (f)) for the top optimal flows found with stretching factor $\lambda = 0.97$ at $Pe = 2^{13}$ and $Pe = 2^{14}$ respectively.}
    \label{fig:OptFlows_Lambda0.97}
\end{figure}

As noted in appendix \ref{AppendixA}, our tests varying the mesh stretching factor $\lambda$ revealed a few cases where the resulting optimal flows slightly outperformed those shown in figure \ref{fig:OptimalWavyFlows}, with moderately improved convergence. 
The two rows of figure \ref{fig:OptFlows_Lambda0.97} show two cases, with $Pe = 2^{13}$ and $2^{14}$, that had noticeable differences in the streamlines compared to those in column A of figure \ref{fig:OptimalWavyFlows} at the same $Pe$. The flows computed with $\lambda = 0.97$ yield $Nu$ values that are $0.62 \%$ and $0.39\%$ higher, respectively, than those obtained with $\lambda = 0.997$. Otherwise, the flows with $\lambda = 0.997$ gave larger Nu, or the differences between the two $\lambda$ were very small, so for simplicity we used $\lambda = 0.997$ for our results in the main text.

%\clearpage
%\newpage
%\bibliographystyle{unsrt}
%\bibliography{OptimalHeatTransfer}

\end{document}